\documentclass[aps,floats]{revtex4}
\usepackage{amsmath}
\usepackage{graphicx,epsfig}

\begin{document}
\bibliographystyle {plain}

\def\oppropto{\mathop{\propto}} 
\def\opsimeq{\mathop{\simeq}}
\def\opoverderline{\mathop{\overline}}
\def\operarrow{\mathop{\longrightarrow}}
\def\opsim{\mathop{\sim}}

\def\fig#1#2{\includegraphics[height=#1]{#2}}
\def\figx#1#2{\includegraphics[width=#1]{#2}}


\title{ Statistics of low energy excitations for the directed
polymer \\
in a $1+d$ random medium ($d=1,2,3$)    }


\author{ C\'ecile Monthus and Thomas Garel }
 \affiliation{Service de Physique Th\'{e}orique, CEA/DSM/SPhT\\
Unit\'e de recherche associ\'ee au CNRS\\
91191 Gif-sur-Yvette cedex, France}

\begin{abstract}
We consider a directed polymer of length $L$ in a random medium of
space dimension $d=1,2,3$. The statistics of low energy excitations
as a function of their size $l$ is numerically evaluated. These
excitations can be divided into bulk and boundary excitations, with
respective densities $\rho^{bulk}_L(E=0,l)$ and $\rho^{boundary}_L(E=0,l)$.
We find that both densities follow the scaling behavior
$\rho^{bulk,boundary}_L(E=0,l) = L^{-1-\theta_d}
R^{bulk,boundary}(x=l/L)$, where $\theta_d$ 
is the exponent governing the energy fluctuations at zero temperature
(with the well-known exact value $\theta_1=1/3$ in one dimension).
In the limit $x=l/L \to 0$, both scaling functions
$R^{bulk}(x)$ and $R^{boundary}(x)$ behave as
$R^{bulk,boundary}(x) \sim x^{-1-\theta_d}$, leading to the droplet
power law $\rho^{bulk,boundary}_L(E=0,l)\sim l^{-1-\theta_d} $ in the
regime $1 \ll l \ll L$. Beyond
their common singularity near $x \to 0$, the two scaling functions 
$R^{bulk,boundary}(x)$ are very different :
whereas $R^{bulk}(x)$ decays monotonically for $0<x<1$, the function
$R^{boundary}(x)$ first decays for $0<x<x_{min}$, then grows for
$x_{min}<x<1$, and finally presents a power law singularity
$R^{boundary}(x)\sim (1-x)^{-\sigma_d}$ near $x \to 1$.
The density of excitations of length $l=L$ accordingly decays as
$\rho^{boundary}_L(E=0,l=L)\sim L^{- \lambda_d} $ where
$\lambda_d=1+\theta_d-\sigma_d$. We obtain $\lambda_1 \simeq
0.67$, $\lambda_2 \simeq 0.53$ and $\lambda_3 \simeq 0.39$, suggesting
the possible relation $\lambda_d= 2 \theta_d$.

\bigskip


\end{abstract}

\maketitle

\section{Introduction }

The model of a directed polymer in a random medium has attracted a lot of
attention in the last twenty years for two main reasons.  On the one
hand, it is directly related to non-equilibrium growth models
\cite{Hal_Zha}; on the other hand, it plays the role of a `baby spin
glass' model in the field of disordered systems
\cite{Hal_Zha,Der_Spo,Der,Mez,Fis_Hus}. This model presents a low
temperature disorder dominated phase, where the 
order parameter is an `overlap' \cite{Der_Spo,Mez,Car_Hu,Com}.
This low temperature phase displays an extreme sensitivity with
respect to temperature or disorder changes  \cite{Fis_Hus,
Zha,Fei_Vin,Sha,Sal_Yos,Sil_Bou,LeDou}, and aging properties for the
dynamics \cite{Yos}.
In finite dimensions, a scaling droplet theory was proposed
 \cite{Fis_Hus,Hwa_Fis},
in direct correspondence with the droplet
 theory of spin glasses \cite{Fis_Hus_SG},
whereas in the mean-field version of the model on the Cayley,
a freezing transition very similar to the one occurring
in the Random Energy Model was found \cite{Der_Spo}.
The phase diagram as a function of space dimension $d$ is the
following \cite{Hal_Zha} : for $d>2$, 
there exists a phase transition between
the low temperature disorder-dominated phase
and a free phase at high temperature  \cite{Imb_Spe,Coo_Der}.
 This phase transition has been studied numerically in $d=3$
\cite{Der_Gol,Kim_Bra_Moo}, exactly on a Cayley tree \cite{Der_Spo}
and on hierarchical lattice \cite{Der_Gri}.
 On the contrary, in dimension $d \leq 2$, there is no free phase,
i.e. any initial disorder drives the polymer into the strong disorder phase.
In this paper, we will be interested in the low energy excitations
above the ground state in dimensions $d=1,2,3$. 

In disordered systems, there can be states that have an energy
very close to the ground state energy but which are very different
from the ground state in configuration space.
For spin glasses, the debate between the droplet
and replica theories concerns the
probabilities and the properties of these states. In the droplet theory
\cite{Fis_Hus_SG}, the low temperature physics is described
in terms of rare regions with nearly degenerate excitations
which appear with a probability that decays with a power law of their size.
In the replica theory \cite{replica}, the replica symmetry breaking
is interpreted as the presence of many pure states
 in the thermodynamic limit,
i.e. the nearly degenerate ground states appear with a finite probability
for arbitrary large size.
More generally, the statistical properties of the nearly degenerate 
excitations (their numbers, their sizes, their geometric properties, 
the barriers separating them, etc...)
are interesting in any disordered system, since they govern
all properties at very low temperature.
In particular, a linear behavior in temperature of the specific heat
$C(T) = b T +O(T^2)$ seems rather generic for a large class of
disordered models, including (i) spin glasses where this behavior is
measured experimentally \cite{binderyoung} and numerically \cite{jerome04}
(ii) disordered elastic systems \cite{schehr} (iii) one-dimensional 
spin models where this behavior can be exactly computed via the
Dyson-Schmidt method \cite{livreluck}.
For the last case, the coefficient $b$ of the linear term of
the specific heat can be put in direct correspondence with the density
$\rho(E=0,l)$ of two-level low energy excitations of size $l$
\cite{us_toy,twolevel,chenma} via the simple formula $ b=(\pi^2/6)
\int dl \rho(E=0,l) $. Other integrals like $\int dl \ l^k
\rho(E=0,l)$ with $k=1,2,...$ determine the low temperature behavior
of other observables.
The explicit computation of the density $\rho(E=0,l)$
of excitations as a function of their size $l$ has been possible
only for one-dimensional models, such as the case of one particle in
a random potentials  \cite{us_toy,twolevel} and the random field Ising chain 
via strong disorder renormalization \cite{twolevel}.
For higher dimensional disordered systems, the statistics of
excitations can only be studied numerically.
In particular, there has been a lot of efforts to characterize the
distribution and the topology of the low energy excitations in
of spin glasses \cite{numerical}. For the directed 
polymer model in finite dimensions, we are only aware
 of the work of Tang \cite{Tang}, where the probability 
of two degenerate non overlapping ground states with binary disorder
in $1+1$ was found to decay as $L^{-2/3}$. Our aim in this paper is to
measure the statistics $\rho_L(E=0,l)$ of low energy excitations of
length $l$, for a directed polymer of length $L$ in a Gaussian random
potential, in dimensions $d=1,2,3$. We compare our results with
the droplet scaling theory in finite dimensions
\cite{Fis_Hus,Hwa_Fis}. We also try to make the connection with
the exact results \cite{Der_Spo} on the Cayley tree ($d=\infty$). 

The paper is organized as follows. In section \ref{previouswork}, we
briefly recall known results on the directed polymer in a random
medium, concerning exact results in $d=1$ and $d=\infty$ (Cayley
tree), as well as the spin glass inspired droplet scaling theory. 
After presenting the parameters of our numerical study in section
\ref{numerics}, we briefly present our results on ground state
energies in section \ref{groundstate}, before focusing on low energy
excitations. The statistics of boundary and bulk
excitations are respectively studied in sections \ref{statexbo} and
\ref{statexbu}. We finally summarize and discuss our results in
section \ref{conclusion}.

\section{ Summary of previous work on directed polymers at low temperature }
\label{previouswork}

\subsection{ Ground state properties }

The probability distribution of the ground state energy $E_0$
of a directed polymer of length $L$ in dimension $1+d$
is expected to follow a scaling form
\begin{eqnarray}
P_d(E_0,L)  \sim \frac{1}{L^{\theta_d}} {\cal P}_d 
\left( {\cal E} = \frac{E_0- L e_0}{L^{\theta_d}} \right)
\label{distrie0}
\end{eqnarray}
where $e_0$ represent the ground state energy density per monomer.
The exponent $\theta_d$ then governs both
the fluctuation and the correction to extensivity of the mean value
(Note that this is not always the case in disordered systems,
see e.g. \cite{Bou_Krz_Mar}).
This result has been proven in $d=1$ with the exact value
of the exponent \cite{Hus_Hen_Fis,Kar,Joh,Pra_Spo}
\begin{eqnarray}
\theta_1=1/3
\label{omegad1}
\end{eqnarray}
For the mean-field version on the Cayley tree ($d=\infty$)
one has formally 
 $\theta_{\infty}=0$ \cite{Der_Spo} :
the width is of order $O(1)$,
whereas the correction to the extensive term $Le_0$ in the averaged value
 $\overline{E_0(L)} $ is of order $O(\ln L)$ \cite{Der_Spo}.
In finite dimensions $d=2,3,4,5,...$, 
the exponent $\theta_d$ has been numerically measured
\cite{Tan_For_Wol,Ala_etal,KimetAla,Mar_etal}.
The values of the KPZ exponent $\chi_d$ measured
 in \cite{Mar_etal} for dimensions $d=2,3$
translate into the following values for the 
directed polymer exponent $\theta_d$
\begin{eqnarray}
\theta_2 && = 0.244 \\
\theta_3 && = 0.186
\label{omegad2et3}
\end{eqnarray}
through the correspondence $\theta_d=\chi_d/(2-\chi_d)$.
Note that the existence of a finite upper critical dimension
has remained a very controversial issue between the numerical studies
\cite{Tan_For_Wol,Ala_etal,KimetAla,Mar_etal}
and various theoretical approaches \cite{Las_Kin,Col_Moo,LeDou_Wie}.
Beyond the exponent $\theta$, 
the scaling function ${\cal P}_d$ itself is also of interest :
it is exactly known
in $d=1$ \cite{Joh,Pra_Spo} (as well as in other geometries
\cite{Bru_Der}), and has been studied on the Cayley tree,
with the conclusion that the distribution is not universal
but depends on the disorder distribution \cite{Dea_Maj}.
Another important property of the ground state
in finite dimensions $d$
is the probability distribution of its end-point 
position $ \vec R_0 $
that follows the scaling form
\begin{eqnarray}
Q_d(R_0,L)  \sim \frac{1}{ \bigl( L^{\zeta_d} \bigr)^d} {\cal Q}_d 
\left( \vec r_0 = \frac{\vec R_0}{L^{\zeta_d}} \right)
\label{distrir0}
\end{eqnarray}
where the exponent $\zeta_d$ is directly related to the previous
exponent $\theta_d$ via the simple relation \cite{Hus_Hen}
\begin{eqnarray}
\zeta_d = \frac{1+\theta_d}{2}
\label{zetaomega}
\end{eqnarray}
This corresponds to a superdiffusive behavior
 $\zeta>1/2$ as soon as $\theta>0$,
and in particular in one dimension $\zeta_1=2/3$.
The probability distribution $Q_d(R_0,L)$ has been studied for the case
$d=1$ in \cite{Hal}. 

\subsection{ Exact identities from the statistical tilt symmetry }

In their continuum version, directed polymers in random media belong
to a special class of disordered models for which
exact remarkable identities for thermal fluctuations
 can be derived \cite{identities88}.
The Hamiltonian of these models have a deterministic part which
consists in quadratic interactions and a random part whose statistics
is translation invariant.
For the directed polymer in dimension $1+1$, this so-called
statistical tilt symmetry leads in particular
to the following
identity for the fluctuation of the end point $r_L$
 \cite{Mez,Fis_Hus,Hwa_Fis}
 \begin{eqnarray}
\overline{ < r_L^2> -<r_L>^2 } = T L 
\label{cumulant}
\end{eqnarray}
Other identities giving relations between higher
cumulants can be similarly derived \cite{identities88,Fis_Hus,Hwa_Fis}.

The identity (\ref{cumulant}) is rather surprising at first sight,
since the fluctuations of the end-point are found to independent
of the disorder and to be exactly the same as in the absence of disorder!
The exact result (\ref{cumulant}) is
particularly interesting at very low temperature,
since it predicts a linear behavior in $T$ of the second cumulant,
and thus puts constraints on the statistics of
nearly-degenerate excitations.
In simpler 1D models of one particle in random potentials
 one can explicitly relate \cite{us_toy,twolevel} the linear
behavior of the position fluctuations to
the rare configurations with two nearly degenerate minima $\Delta E \sim T$.
For directed polymer in random media, any theory of low energy excitations
has to reproduce the result (\ref{cumulant}), and we will now
describe its interpretation within the droplet theory 
\cite{Fis_Hus,Hwa_Fis}.

\subsection{ Droplet theory 
for low energy excitations in finite dimensions }

The droplet theory for directed polymers \cite{Fis_Hus,Hwa_Fis},
is very similar to the droplet theory of spin glasses \cite{Fis_Hus_SG}.
It is a scaling theory that can be summarized as follows.
The exponent $\theta$ involved in the fluctuations of
the ground state energy $E_0$ over the samples (see eq. (\ref{distrie0}))
also governs the fluctuations of the energy within one sample
as the end point varies. 
As a consequence, if one assumes a scaling distribution analogous to
 (\ref{distrie0}), the probability to find a nearly degenerate 
ground state $\Delta E \sim T$ of order $L$ behaves
as $T/L^{\theta}$, so it is rare, but it corresponds
to a very large fluctuation of the end-point of order
$\Delta r \sim L^{\zeta}$ with the spatial exponent $\zeta$
 ( see eq. (\ref{distrir0})).
The contribution of these rare nearly degenerate paths to
the disorder averaged fluctuations (\ref{cumulant}) is thus of order
$(T/L^{\theta}) \times ( L^{\zeta})^2 = T L^{ 2 \zeta - \theta} = T L$
using the scaling relation (\ref{zetaomega}).
 
This naive `zero-order' argument has to be refined if
one is interested into the density $\rho_L(E=0,l)$
of excitations that involve an arbitrary length $l$ for a polymer of
length $L$. 
For $1 \ll l \leq L$, the droplet theory assumes that the same scalings
apply : the probability of an excitation of length $l$ is of order
$T/l^{\theta}$ and leads to a fluctuation
of the end-point of order $\Delta r \sim l^{\zeta}$.
However now, the crucial notion of `independent excitations' \cite{Fis_Hus}
has to be introduced to obtain a consistent picture.
Both for spin glasses \cite{Fis_Hus_SG}
and for directed polymers \cite{Fis_Hus}, the idea
is that in a given sample,
droplets with neighbouring sizes tend to have a big overlap.
More precisely for the polymer, two excitations 
of lengths $l_1$ and length $l_2$ with $l_1 \sim l_2$
will typically merge and then follow the same path to join the ground state,
as shown by the tree structure of optimal paths to all end-points
 \cite{Hal_Zha}. As a consequence in \cite{Fis_Hus},
droplets are considered to be independent only if there is a factor
of order $2$ between their sizes, and this
gives a factor $d \ln l = dl/l$ in all integrations over droplets
\begin{eqnarray}
dl \rho^{indep}(E=0,l)  \sim \frac{ dl }{l^{\theta+1}}
\label{rhodroplet}
\end{eqnarray}
A more intuitive view of the $dl/l$ factor is to remark that
independent excitations stem from a branching process along the ground
state path, and that the infinitesimal number of branches between $l$
and $l+dl$ is precisely $dl/l$.

In summary, the droplet theory predicts a power law
distribution of independent excitations, with exponent $(1+\theta)$.
 Note that the absence of any characteristic scale in $l$
means that there exists some infinite correlation length in the system
in the whole low temperature phase.
This is in contrast with simpler models
like the random field Ising chain \cite{twolevel}
or in the spin glass chain in external field \cite{chenma},
where the density of excitations was found to
decay exponentially in $l$, the correlation length being the Imry-Ma length.

\subsection{ Exact results for low energy excitations on the Cayley tree }

Many exact results for directed polymers on the Cayley tree
have been derived by Derrida and Spohn \cite{Der_Spo}. 
From the point of view of excitations, 
the most important result concerns the distribution
of the overlap in the thermodynamic limit, which is simply the sum of
two delta peaks at $q=0$ and $q=1$ in
the whole low temperature phase \cite{Der_Spo} : 
\begin{eqnarray}
\pi (q)= (1-Y) \delta(q) +Y \delta(q-1)
\label{piarbre}
\end{eqnarray}
and the distribution of $Y$ over the samples is 
the same as in the Random Energy Model \cite{Der_REM}.
In particular, the disorder average of eq. (\ref{piarbre}) is
\cite{Der_Spo} :  
\begin{eqnarray}
\overline{\pi (q)}= \frac{T}{T_c} \delta(q)
 + \left( 1- \frac{T}{T_c} \right) \delta(q-1)
\label{overlapcayley}
\end{eqnarray}
i.e. the overlap is zero with probability $T/T_c$, and one otherwise.
This means that for a polymer of length $L$,
the important excitations are those of length $l \sim L$,
and that these excitations keep 
a finite weight in the limit $L \to \infty$.
To understand the origin of this surprising result,
Fisher and Huse \cite{Fis_Hus} have computed that the probability 
to find an excitation of length $l$ which branches off
at a distance $s=L-l \ll L$ from the root behaves as $s^{-3/2}$,
i.e. using the length notation $l=L-s$
\begin{eqnarray}
 \rho_L(E=0,l)  \sim \frac{ 1  }{(L-l)^{3/2}}
\end{eqnarray}
Finally,
Tang has studied numerically the 
overlap distribution $ \overline{P_L(q)}$
for a polymer of finite-size $L$
 \cite{Tang} to characterize how the
two delta peaks develop in (\ref{overlapcayley}) :
the data for $0<q<1$ follow the scaling behavior
(see Figs 4(a) and 4(b) of \cite{Tang})
 \begin{eqnarray}
\overline{\pi_L (q)} \sim L^{- 1/2} {\hat \pi} (q)
\label{overlaptang}
\end{eqnarray}
where the scaling function ${\hat \pi} (q)$ 
present the same singularity with exponent $3/2$
near $q \to 0$ and $q \to 1$
\begin{eqnarray}
{\hat \pi} (q) && \oppropto_{q \to 0} \frac{1}{q^{3/2} } \\
{\hat \pi} (q) && \oppropto_{q \to 1} \frac{1}{(1-q)^{3/2} }
\label{overlaptangsing}
\end{eqnarray}
so that, in the limit $L \to \infty$, $\overline{\pi_{\infty}(q)}$
only contains two $\delta$ peaks at $q=0$ and $q=1$ (see
eq. (\ref{overlapcayley})). 

\section{ Model and numerical details}
\label{numerics}
In this paper, we present numerical results for
the random-bond version of the directed polymer model on a $1+d$
hypercubic lattice. The partition function satisfies the following
recursion
\begin{eqnarray}
Z_{L+1} (\vec r) =  \sum_{j=1}^{2d}
 e^{-\beta \epsilon_L(\vec r+\vec e_j,\vec r)} Z_{L} (\vec r+\vec e_j)
\end{eqnarray}
The bond-energies $\epsilon_L(\vec r+\vec e_j,\vec r) $
are random independent variables, drawn with the Gaussian distribution
\begin{eqnarray}
\rho (\epsilon) = \frac{1}{\sqrt{2\pi} } e^{- \frac{\epsilon^2}{2} }
\end{eqnarray}

For each dimension $d$, we now give the typical lengths $L$ 
we have studied, with the corresponding number $n_s$ of disordered
samples :

(i) For $d=1$,  $L=50,100,200,300,400,600$, with respective
$n_s/10^6=160,40,10,4.7,2.5,4.5$. 

(ii) For $d=2$, $L=20,40,60,80,100$, with respective
$n_s/10^6=30,4.4,5.8,2.2,1$. 

(iii) For $d=3$, $L=12,18,24,30,36,42$, with respective
$n_s/10^6=12.5,2.8,0.9,1.6,0.76,0.4$. 

We first briefly describe our results on the ground state energy
statistics, before we turn to the measure of the density of low energy
excitations on which we focus our attention in this paper.

\section{ Distribution of the ground state energy }
\label{groundstate}
\subsection{ Scaling distribution   }

\begin{figure}[htbp]
\includegraphics[height=6cm]{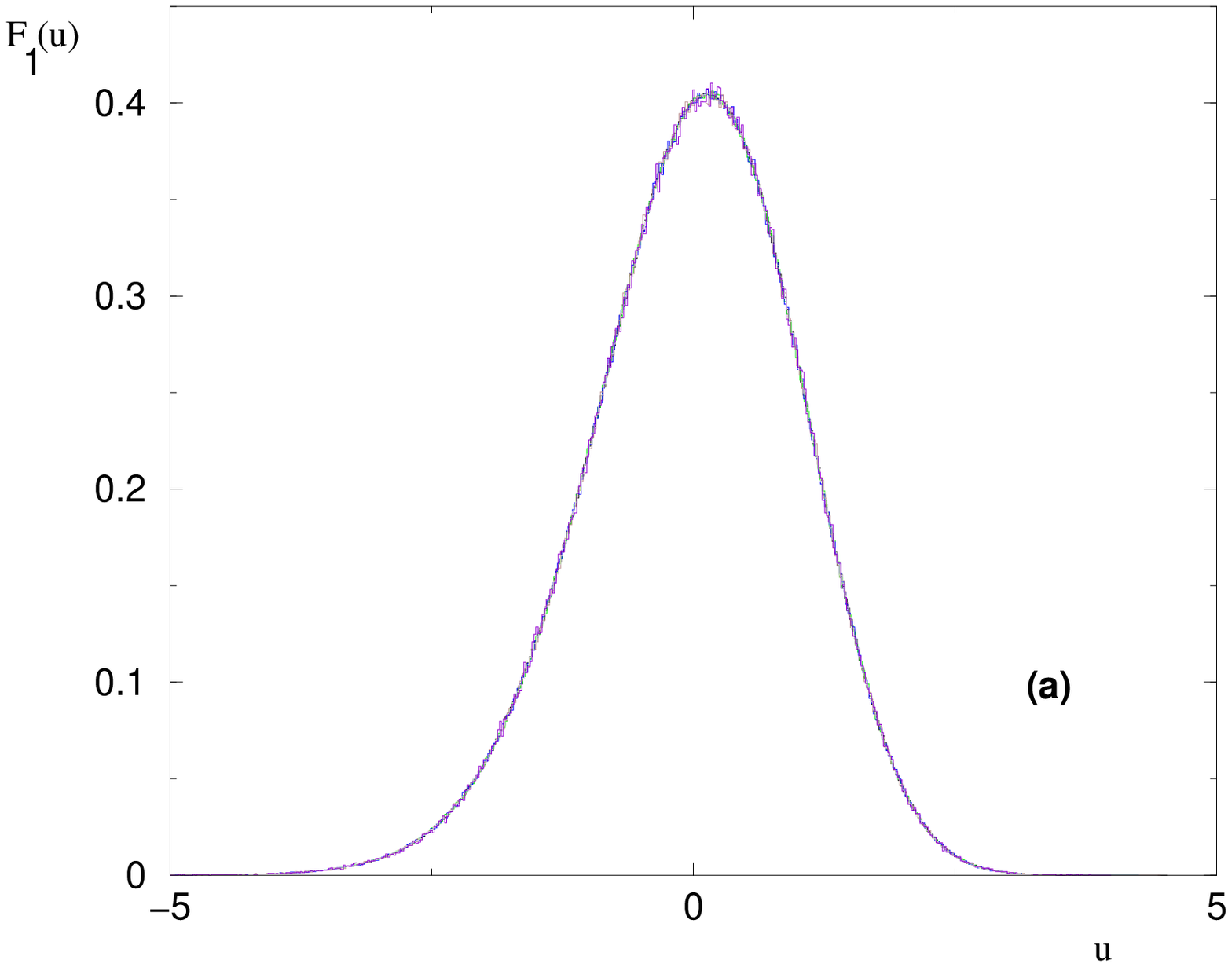}
\hspace{1cm}
\includegraphics[height=6cm]{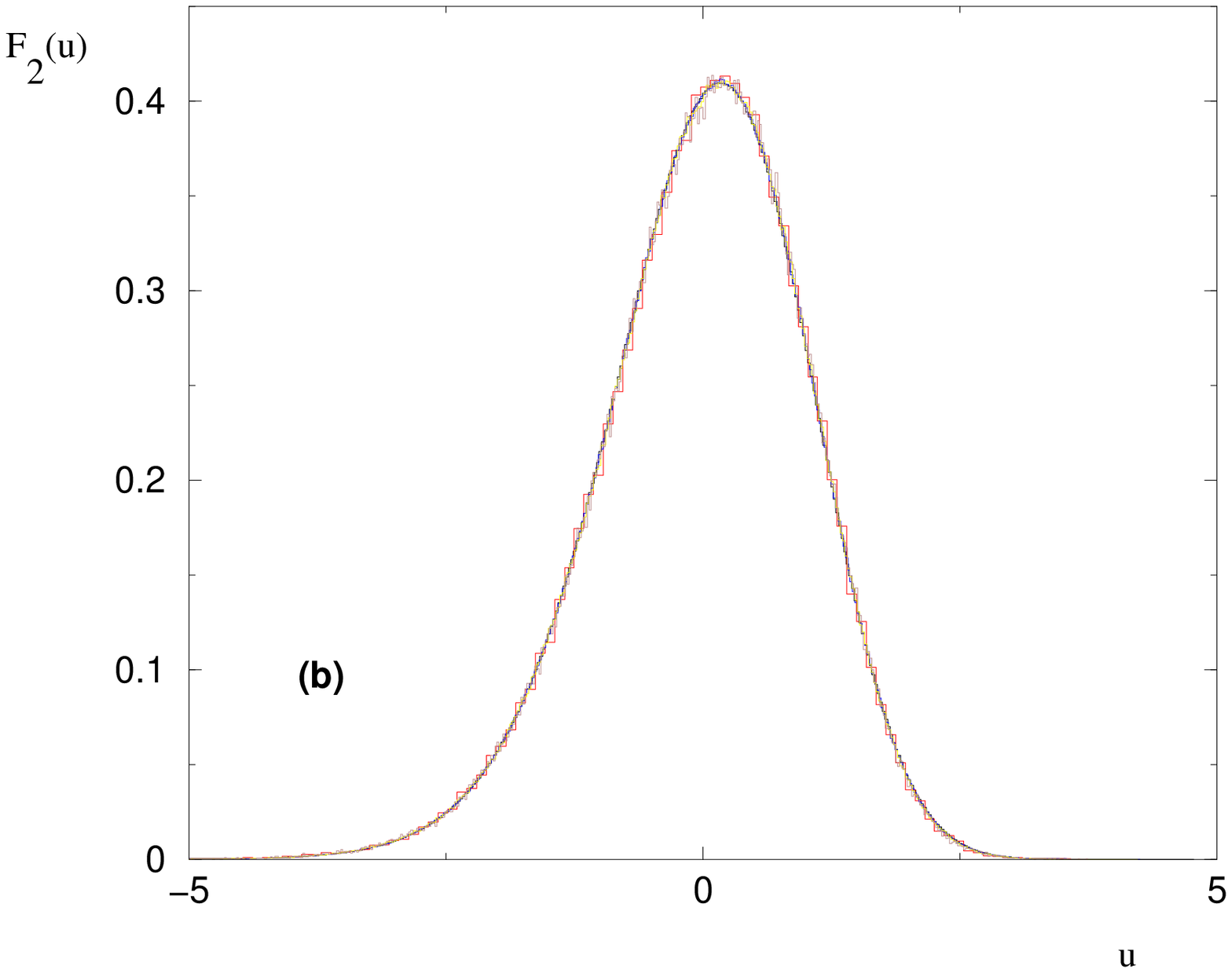}
\caption{(a) Rescaled distribution $F_1(u)$ of the $d=1$ ground state
energy (see equation (\ref{rescalinghistoe0})), for
$L=50,100,200,300,400,600$. 
(b) Rescaled distribution $F_2(u)$ of the $d=2$ ground state energy
(see equation (\ref{rescalinghistoe0})), for $L=10,20,40,80,120$.}
\label{fig1}
\end{figure}

\begin{figure}[htbp]
\includegraphics[height=6cm]{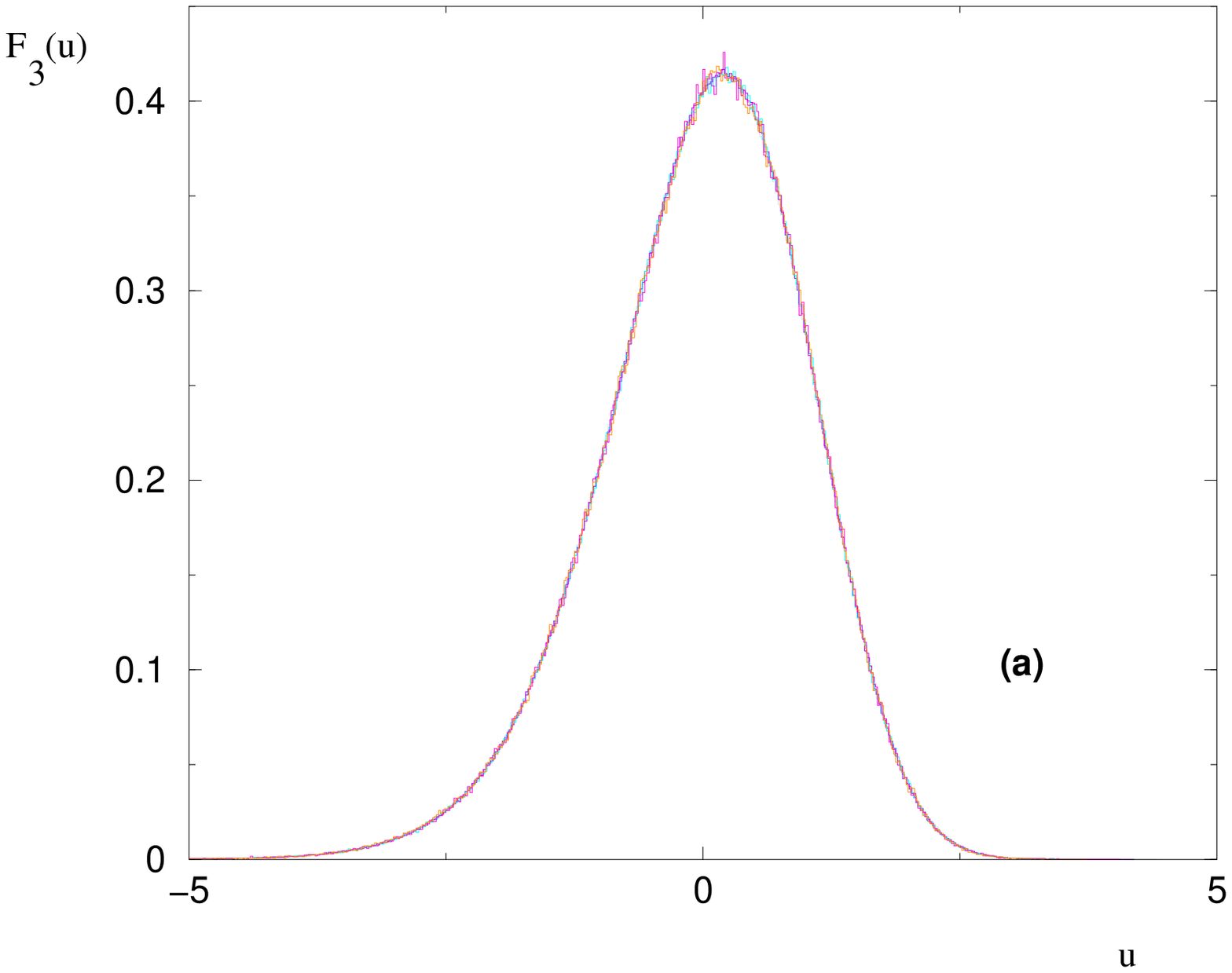}
\includegraphics[height=6cm]{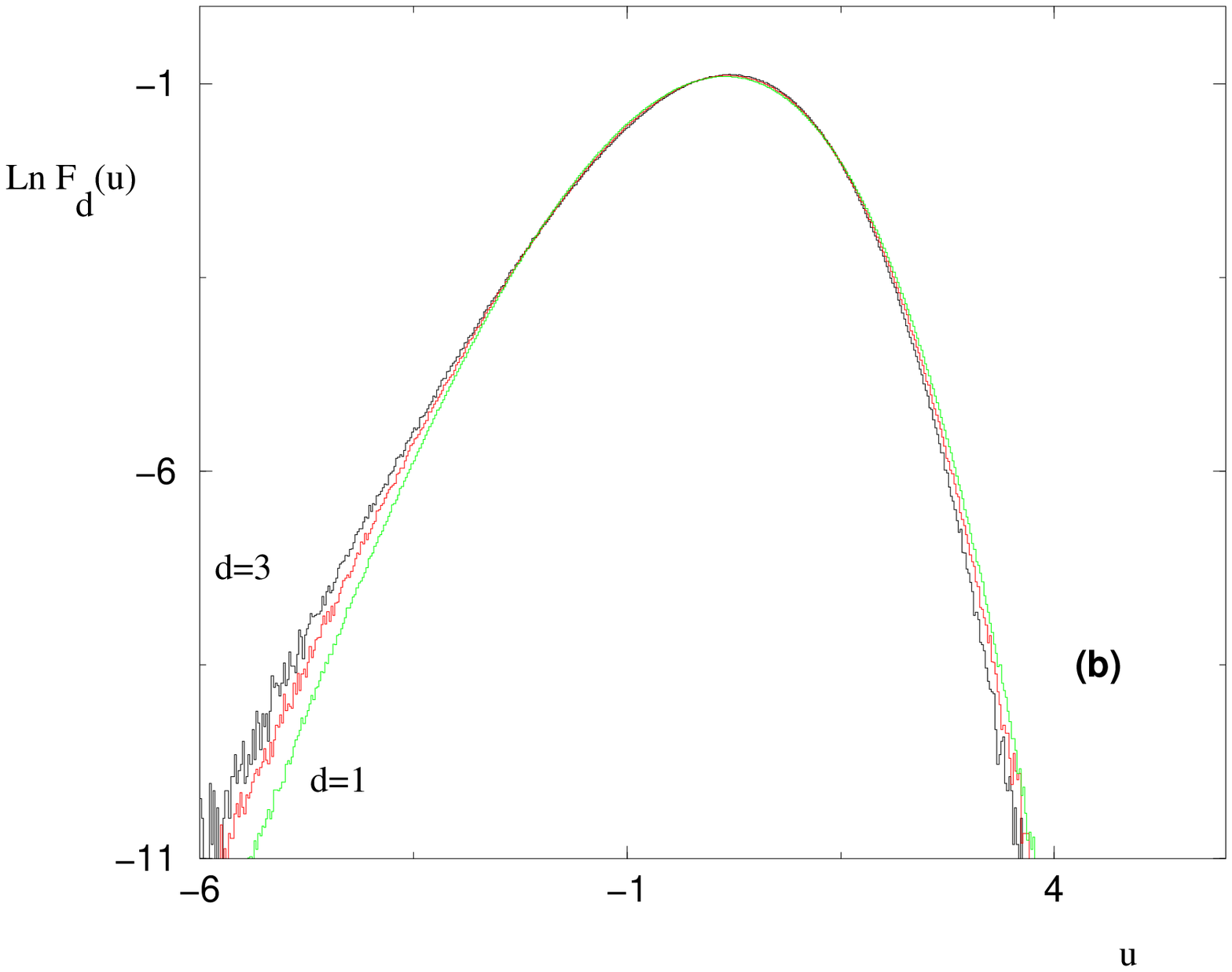}
\caption{(a) Rescaled distribution $F_3(u)$ of the $d=3$ ground state
energy (see equation (\ref{rescalinghistoe0})), for
$L=6,12,18,24,36$. (b) Comparison of ${\rm Ln} \ F_d(u)$ for
$d=1,2,3$ (see equation (\ref{rescalinghistoe0})).}
\label{fig2}
\end{figure}

For the sizes $L$ we have considered,
the distribution of the ground state energy $E_0$
follows the scaling form
\begin{equation}
P_d(E_0,L)   \simeq  \frac{1}{ \Delta E_0(L)} \  
F_d \left( x= \frac{
E_0 -E_0^{av}(L)}{ \Delta E_0(L) }  \right) 
\label{rescalinghistoe0}
\end{equation}
as shown on Figs \ref{fig1}(a), \ref{fig1}(b), \ref{fig2}(a)
for $d=1,2,3$ respectively. 
We have checked that the function $F_1(u)$ of Fig. \ref{fig1}(a)
 agrees with the numerical tabulation of the exact result given 
on the web site \cite{website}.
The three functions $F_1, F_2, F_3$ are
shown together for comparison on Fig. \ref{fig2}(b).

\subsection{ Behavior of the width 
$\Delta E_0(L)$ and average $E_0^{av}(L)$}

The exponent $\theta_d$ of eq.(\ref{distrie0})
is expected to govern both the width $\Delta E_0(L)$
and the correction to extensivity of the average
\begin{eqnarray}
\Delta E_0(L) && \sim  L^{\theta_d} \\
\frac{E_0^{av}(L)}{L} && \sim e_0(d) + L^{\theta_d-1} e_1(d) +...
\end{eqnarray}

Our measures of the exponent $\theta_d$ from the 
width $\Delta E_0(L)$ yield
\begin{eqnarray}
\theta_1 && \sim 0.33 \\
\theta_2 && \sim 0.24 \\
\theta_3 && \sim 0.18
\end{eqnarray}
are in agreement with the exact value 
$\theta_1=1/3$ and with the previous
numerical measures quoted in Eq. (\ref{omegad2et3}) for $d=2,3$.

The fits of the average yields
\begin{eqnarray}
d=1 : \ \ \ \frac{E_0^{av}(L)}{L} && \sim -0.95 +
0.84  \ L^{\theta_1-1} - 1.19 \ L^{-1} \\
d=2 : \ \ \ \frac{E_0^{av}(L)}{L} && \sim -1.53 +
 1.48 \ L^{\theta_2-1} - 0.94 \ L^{-1} \\
d=3 : \ \ \ \frac{E_0^{av}(L)}{L} && \sim -1.81 +
 2.05 \ L^{\theta_3-1} - 1.43 \ L^{-1} 
\end{eqnarray}

\section{Statistics of boundary excitations }
\label{statexbo}

\subsection{ Measure of independent excitations}

\begin{figure}[htbp]
\includegraphics[height=8cm]{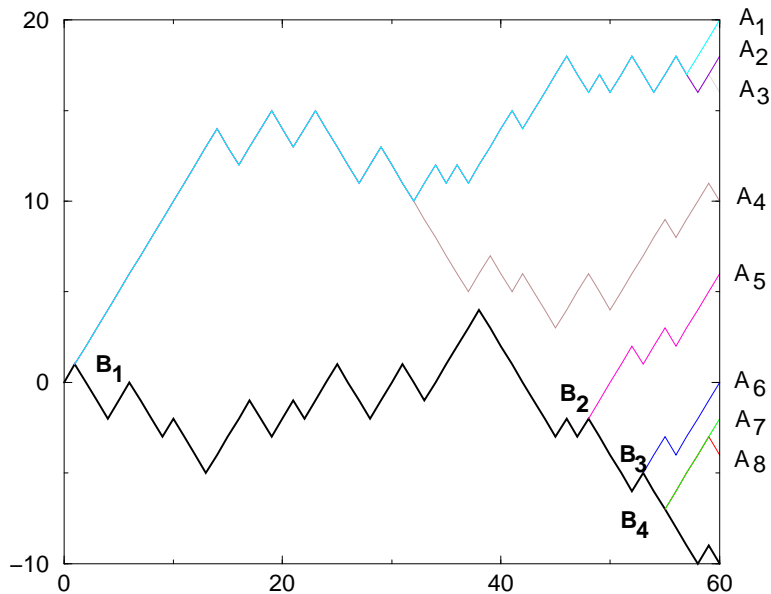}
\caption{ Notion of independent boundary excitations (in $d=1$) :
for each sample, we generate the ground state (bold line) of energy $E_0$,
and consider all end-points $A_i$ where the energy $E(A_i)$
of the best path ending at $A_i$ satisfies $E(A_i)-E_0<T$ (light lines).
These best paths tend to cluster into families.
In our counting procedure of independent excitations,
 two excitations are independent if they have no bond in common.
For instance on the Figure, end-points $A_1$, $A_2$, $A_3$ and $A_4$
count for a single excitation associated to the branch point $B_1$.
Similarly $A_7$ and $A_8$ count for a single excitation
 associated to the branch point $B_4$.
}
\label{figexbo}
\end{figure}

In this Section, we are interested into the density
 $\rho^{boundary}_L(E=0,l)$
of boundary excitations defined as
\begin{eqnarray}
\rho^{boundary}_L(E=0,l) = \lim_{T \to 0} \left[
 \frac{1}{T} \overline{ {\cal N}^{boundary}_L(E<T,l) }  \right]
\label{defrhoboundary}
\end{eqnarray}
where $\overline{{\cal N}^{boundary}_L(E<T,l)}$ 
is the disorder averaged number of independent boundary excitations
of energy $0<E<T \to 0$ and of length $l$ existing for a directed
polymer of length $L$. This number ${\cal N}^{indep}_L(E<T,l)$
is measured as follows.
For each sample, we consider all end-points $\vec r$
different from the ground state $\vec r_0$ : 
if the energy $E_{min}(\vec r)$
of the best path ending at $\vec r$ satisfies $E_{min}(\vec r)-E_0<T$,
we construct the best path ending at $\vec r$ to measure its length $l$,
i.e. the length over which it is different from the ground state.
However, as explained already above when summarizing
the droplet theory (\ref{rhodroplet}), one is interested into
the number of independent excitations. 
So here we have used the following criterion : 
two excitations are independent if they have no bond in common.
In $d=1$, since we use polymers with $(\pm)$ steps, this means
that two excitations are independent if they join the ground state
at different points (see Fig. \ref{figexbo}). In $d=2$ and $d=3$,
two excitations are allowed to join the ground state at the same point
provided they branch along two different directions.

In (\ref{defrhoboundary}), the parameter $T$ is simply a cut-off
and we have checked the independence of
the density $\rho^{boundary}(E=0,l) $
 with respect to $T$ for $T$ small enough.
For instance in $d=2$ and $d=3$, we have checked that 
the cut-offs $T=0.1$ and $T=0.05$
yield the same density  $\rho^{boundary}(E=0,l) $.

\subsection{ Boundary excitations for $d=1$ }

\begin{figure}[htbp]
\includegraphics[height=6cm]{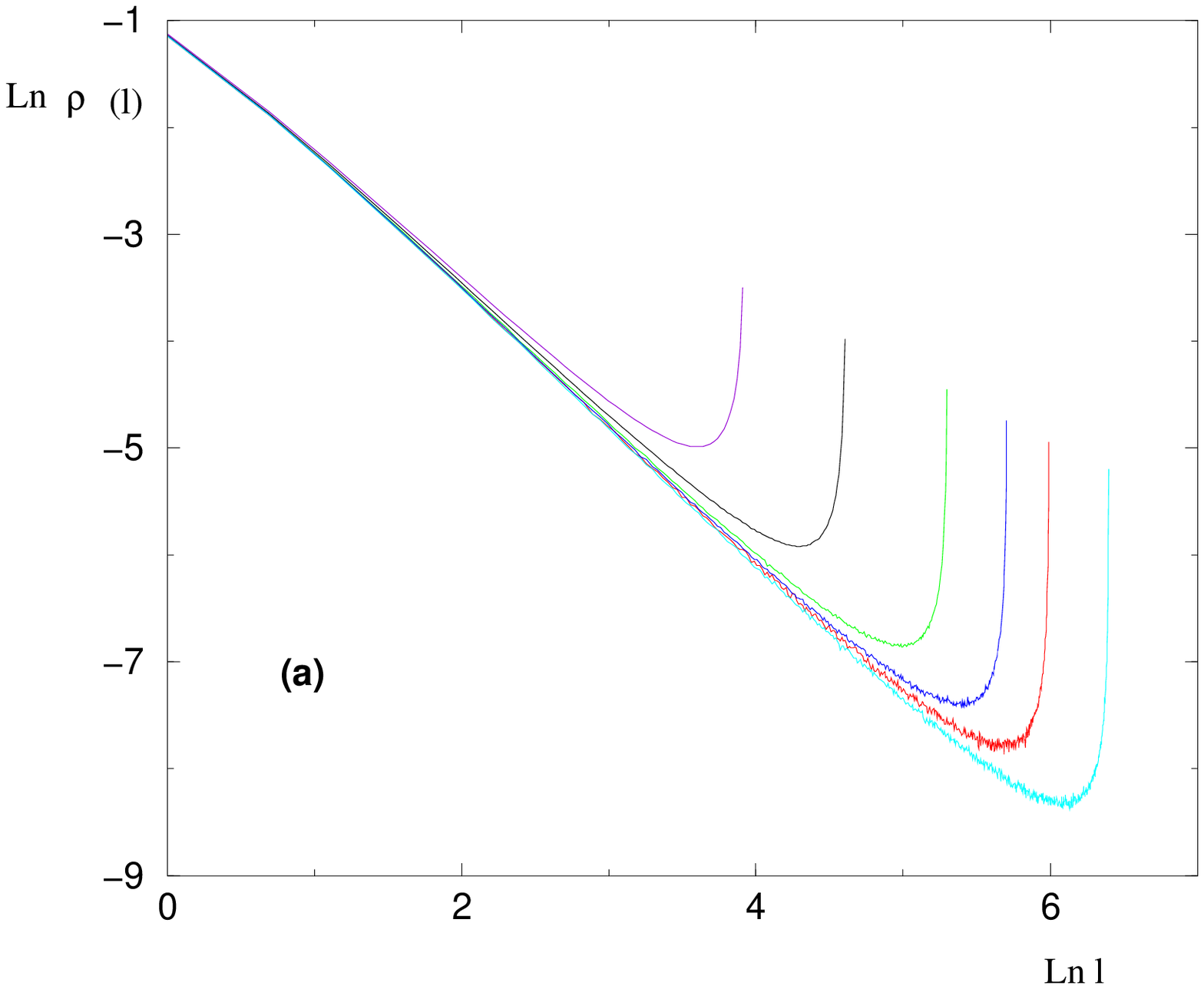}
\hspace{1cm}
\includegraphics[height=6cm]{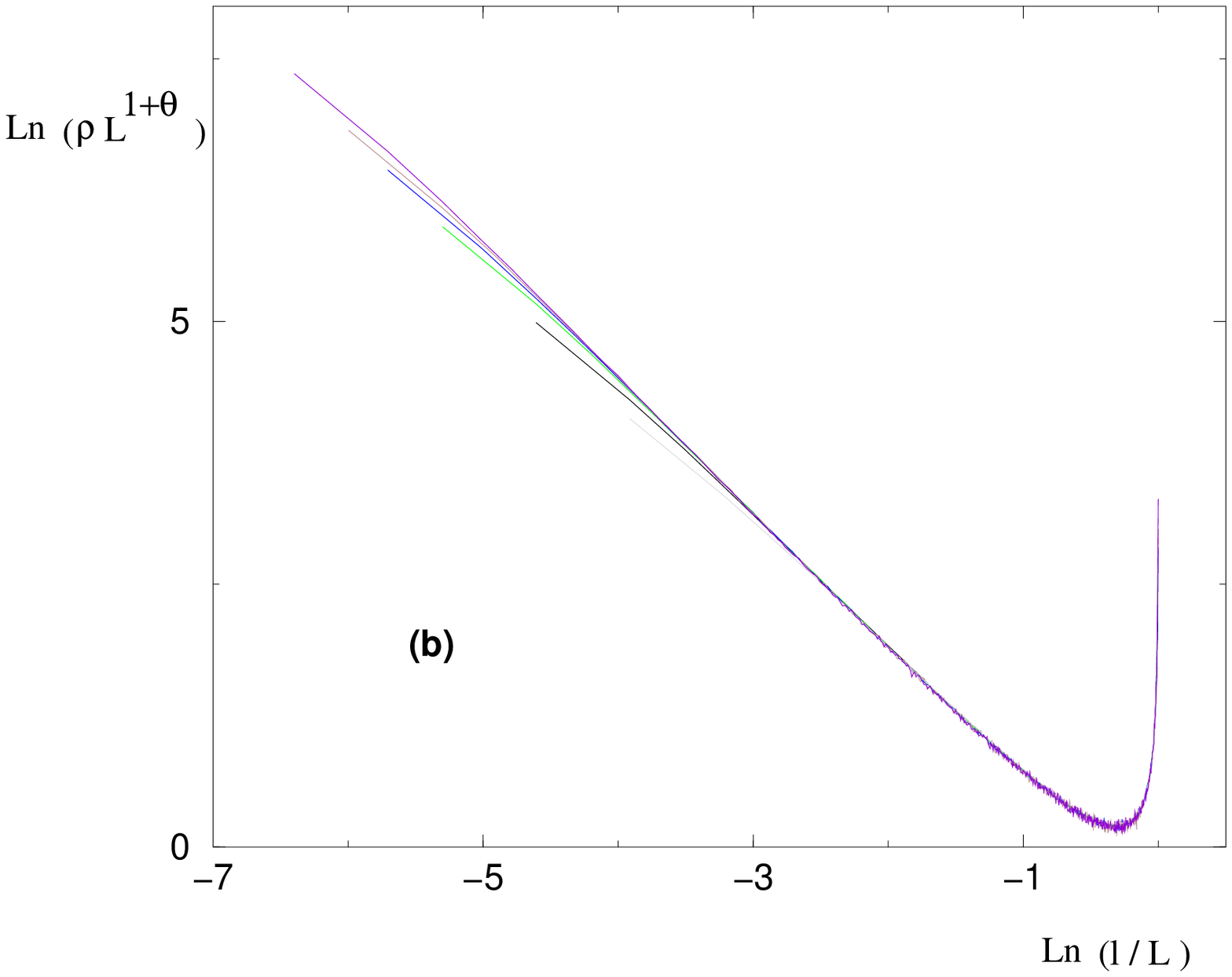}
\caption{(a) Log-log plot of the density $\rho^{boundary}_L(E=0,l)$
of boundary excitations of length $l$ in $d=1$ for
$L=50,100,200,300,400,600$.
(b) Same data after the rescaling of equation (\ref{scalingboundary})
with
$\theta_1=1/3$.}
\label{fig3}
\end{figure}

\begin{figure}[htbp]
\includegraphics[height=6cm]{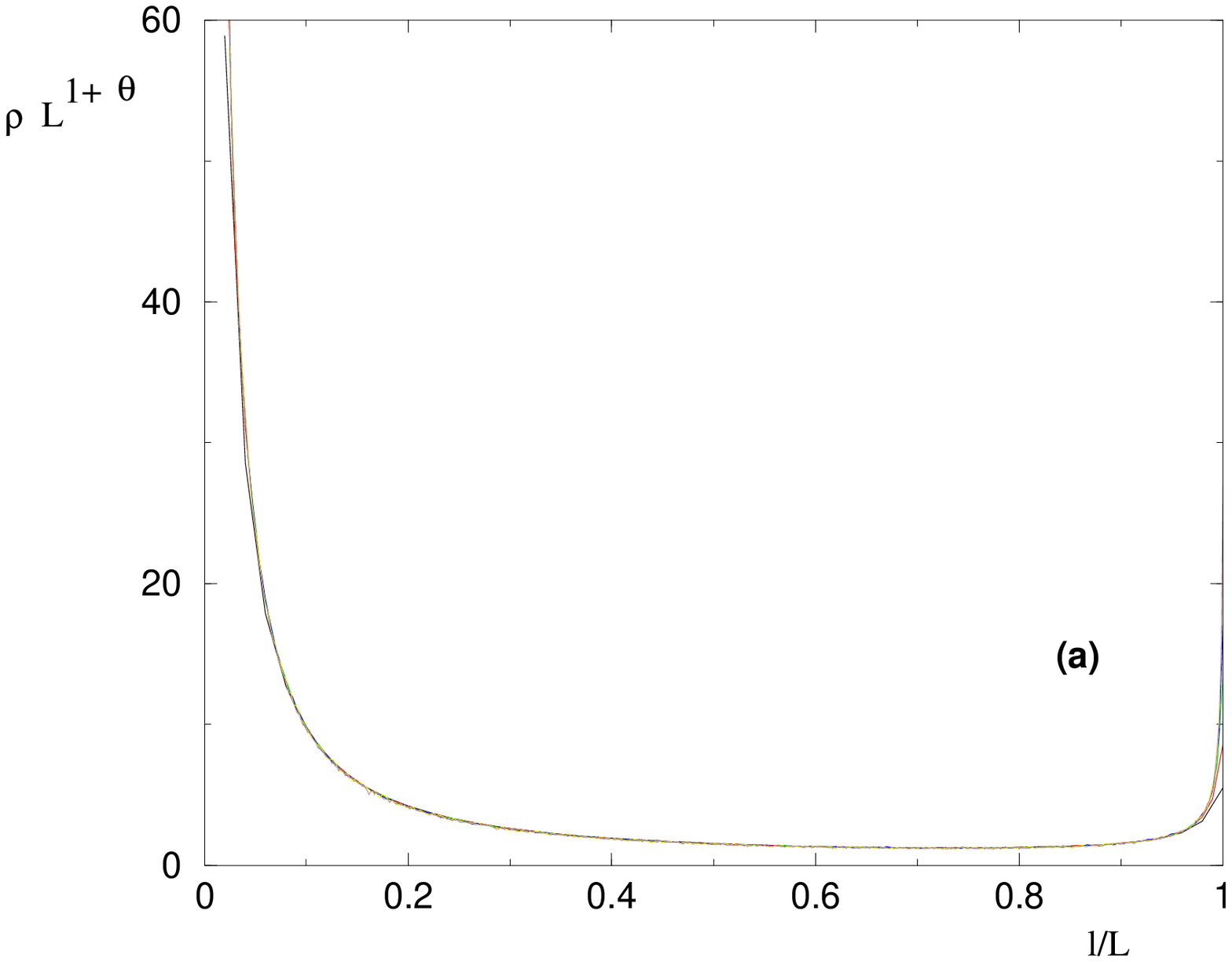}
\hspace{1cm}
\includegraphics[height=6cm]{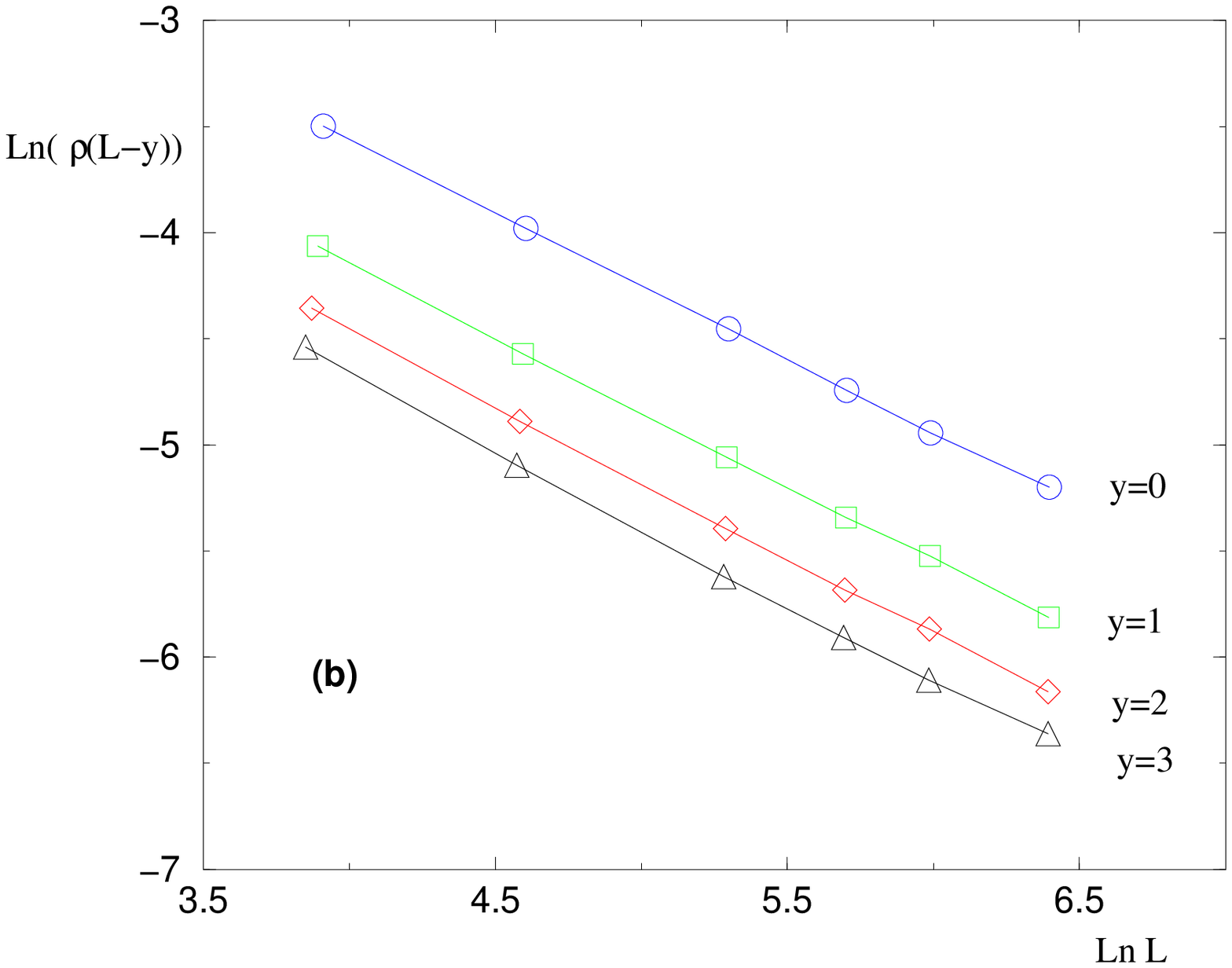}
\caption{$d=1$: (a) Rescaled distribution $R^{boundary}(x=l/L)$
(see equation (\ref{scalingboundary})), that is singular for $x \to 0$
and $x \to 1$.
(b) Log-log plot of the density $\rho^{boundary}_L(E=0,l=L-y)$
of very large boundary excitations $l=L-y$ with finite $y=0,1,2,3$.}
\label{fig4}
\end{figure}

On Fig. \ref{fig3}(a), the density $\rho^{boundary}_L(E=0,l)$
is shown in a log-log plot for various sizes $L$.
These curves can be rescaled according to 
\begin{eqnarray}
\rho^{boundary}(E=0,l) = \frac{1}{L^{1+\theta_1}}
 R^{boundary} \left( x = \frac{l}{L} \right) 
\label{scalingboundary}
\end{eqnarray}
as shown on Fig. \ref{fig3}(b). 
The master curve $R^{boundary}(x)$ shown on Fig. \ref{fig4}(a)
has three important properties 

(i) In the region $x \to 0$, 
the scaling function $R^{boundary}(x)$ follows the power law
(see the log-log plot on Fig. \ref{fig3}(b))
\begin{eqnarray}
R^{boundary}(x) \oppropto_{x \to 0} \frac{1}{x^{1+\theta_1}}
\end{eqnarray} 
so that in the regime $ 1 \ll l \ll L$, 
the statistics of independent excitations
\begin{eqnarray}
\rho^{boundary}_L(E=0,l) \sim \frac{1}{l^{1+\theta_1}}
\ \ \ \hbox{for} \ \ \  1 \ll l \ll L
\end{eqnarray}
follows the droplet power law (\ref{rhodroplet}).

(ii) The function $R^{boundary}(x)$ is minimum at some finite value $x_{min}
(d=1) \sim 0.74$, and then grows for $x_{min}<x<1$

(iii) In the regime $x=l/L \to 1$, the scaling function $R^{boundary}(x)$
diverges. 
 To describe the regime of these very large excitations,
let us first consider the extreme case $l=L$
of an excitation that branches off at the origin. Our result for $l=L$
follow the scaling behavior 
 \begin{eqnarray}
\rho^{boundary}_L(E=0,l=L) \sim \frac{1}{L^{\lambda_1}}
\ \ \ \hbox{with} \ \ \  \lambda_1 \sim 0.67
\label{lambda1d}
\end{eqnarray}
in agreement with the exponent measured by Tang \cite{Tang}
for the probability of two degenerate ground states 
in the case of binary disorder.
More generally, we find that $\rho^{boundary}_L(E=0,l=L-y)$
with finite $y$ also decays as in (\ref{lambda1d}),
see Fig. \ref{fig4}(b).
Since the exponent $\lambda_1 \sim 0.66$ is different from 
the exponent $1+\theta_1=4/3$ appearing in the scaling form 
 (\ref{scalingboundary}), the  singularity of the scaling function
$R^{boundary}(x)$ near $x \to 1$ is given by
\begin{eqnarray}
R^{boundary}(x) \oppropto_{x \to 1} \frac{1}{(1-x)^{\sigma_1}}
\ \ \ \hbox{with} \ \ \  \sigma_1 = 1+\theta_1-\lambda_1 \sim 0.66
\end{eqnarray}

\subsection{ Boundary excitations for $d=2$ }

\begin{figure}[htbp]
\includegraphics[height=6cm]{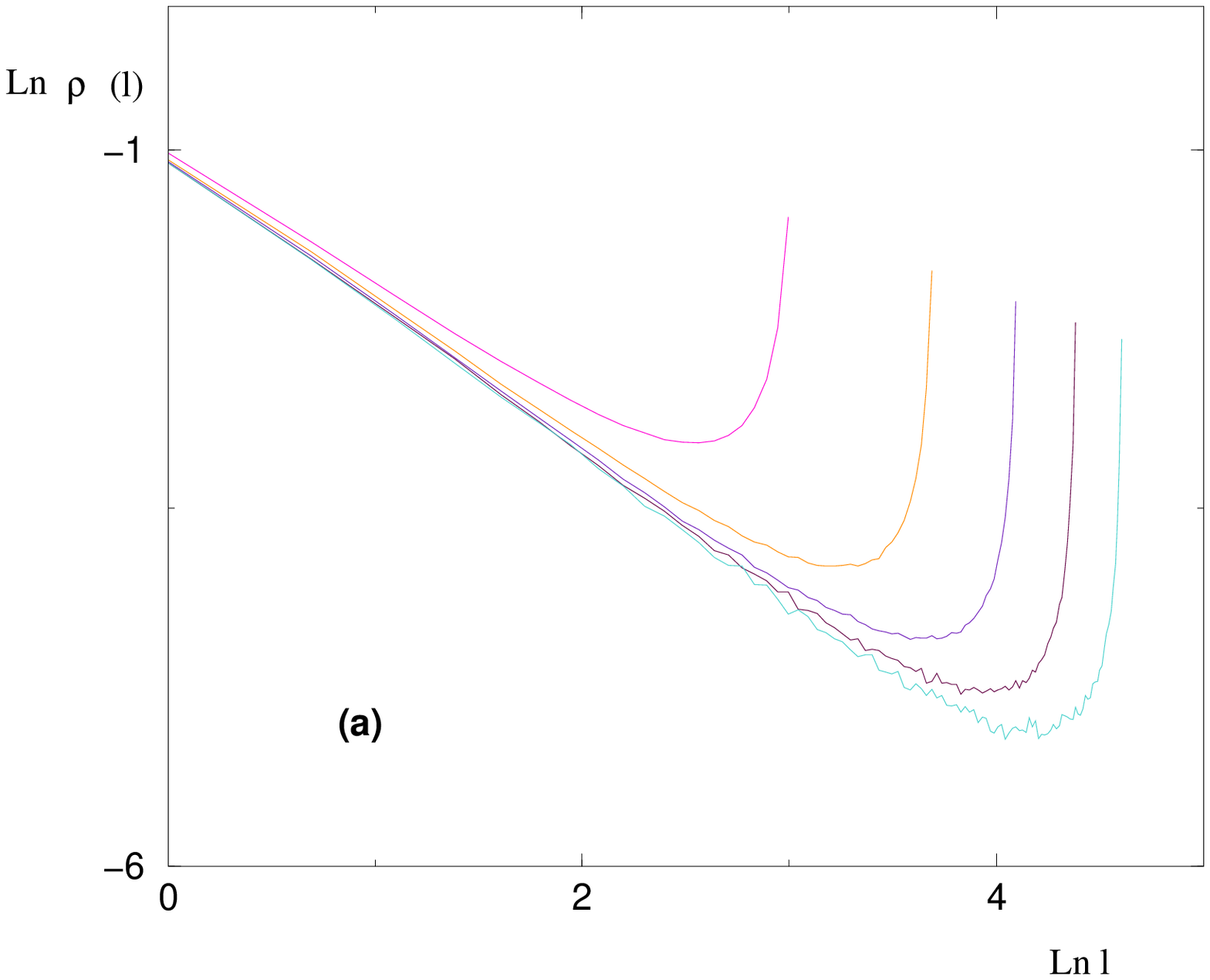}
\hspace{1cm}
\includegraphics[height=6cm]{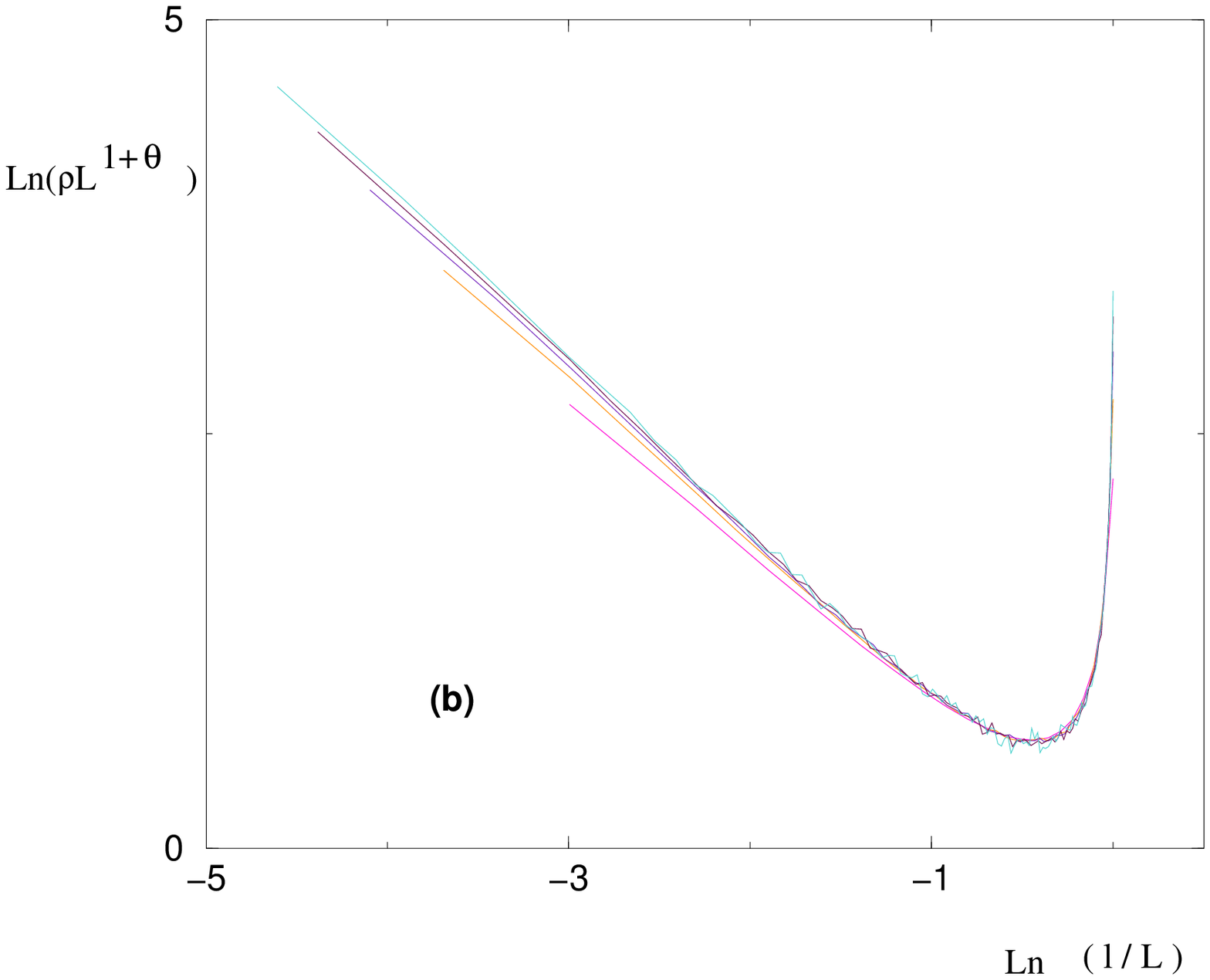}
\caption{(a) Log-log plot of the density $\rho^{boundary}_L(E=0,l)$
of boundary excitations of length $l$ in $d=2$ for
$L=20,40,60,80,100$.
(b) Same data after the rescaling of equation (\ref{scalingboundary2})
with $\theta_2 \sim 0.24$.}
\label{fig5}
\end{figure}

\begin{figure}[htbp]
\includegraphics[height=6cm]{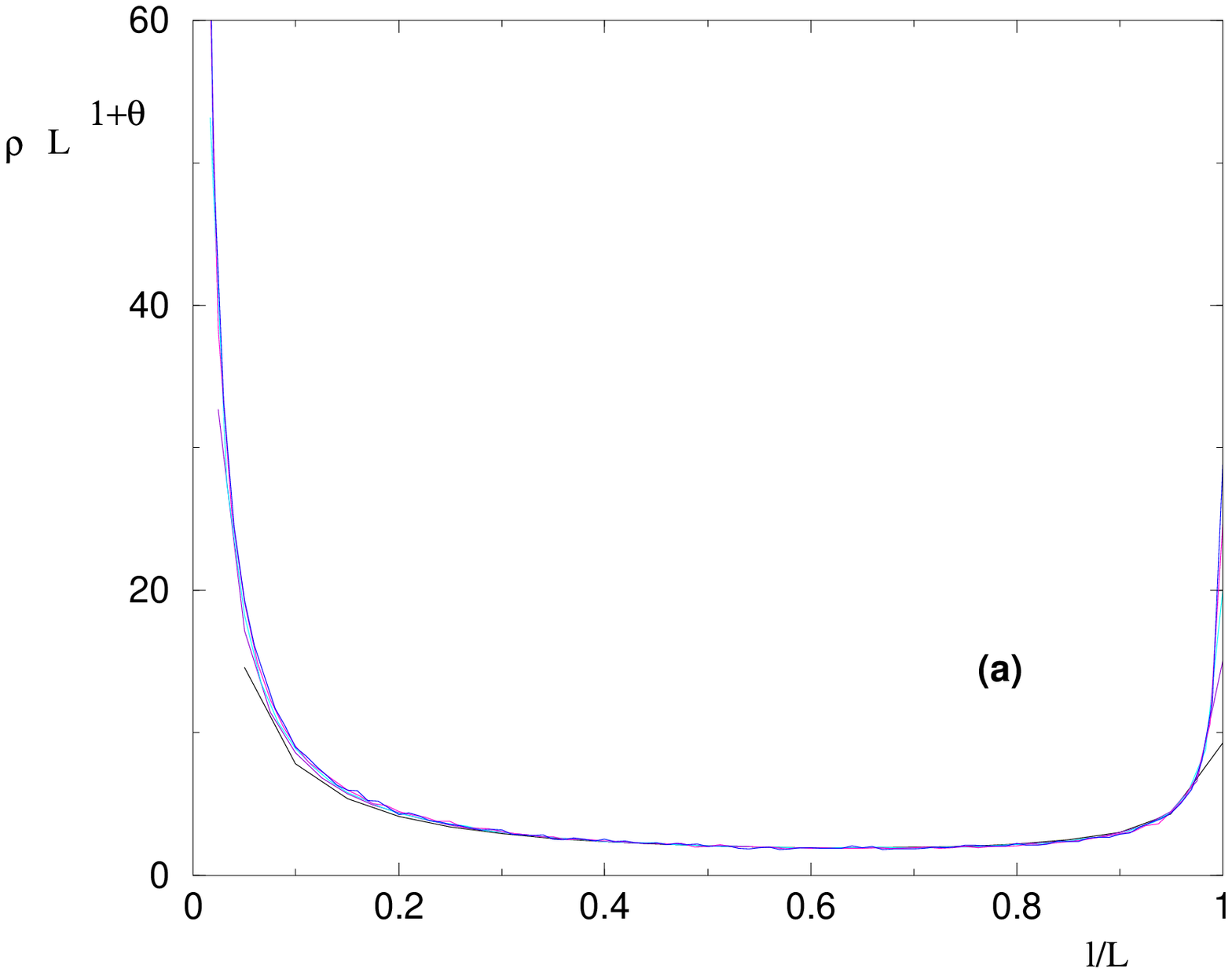}
\hspace{1cm}
\includegraphics[height=6cm]{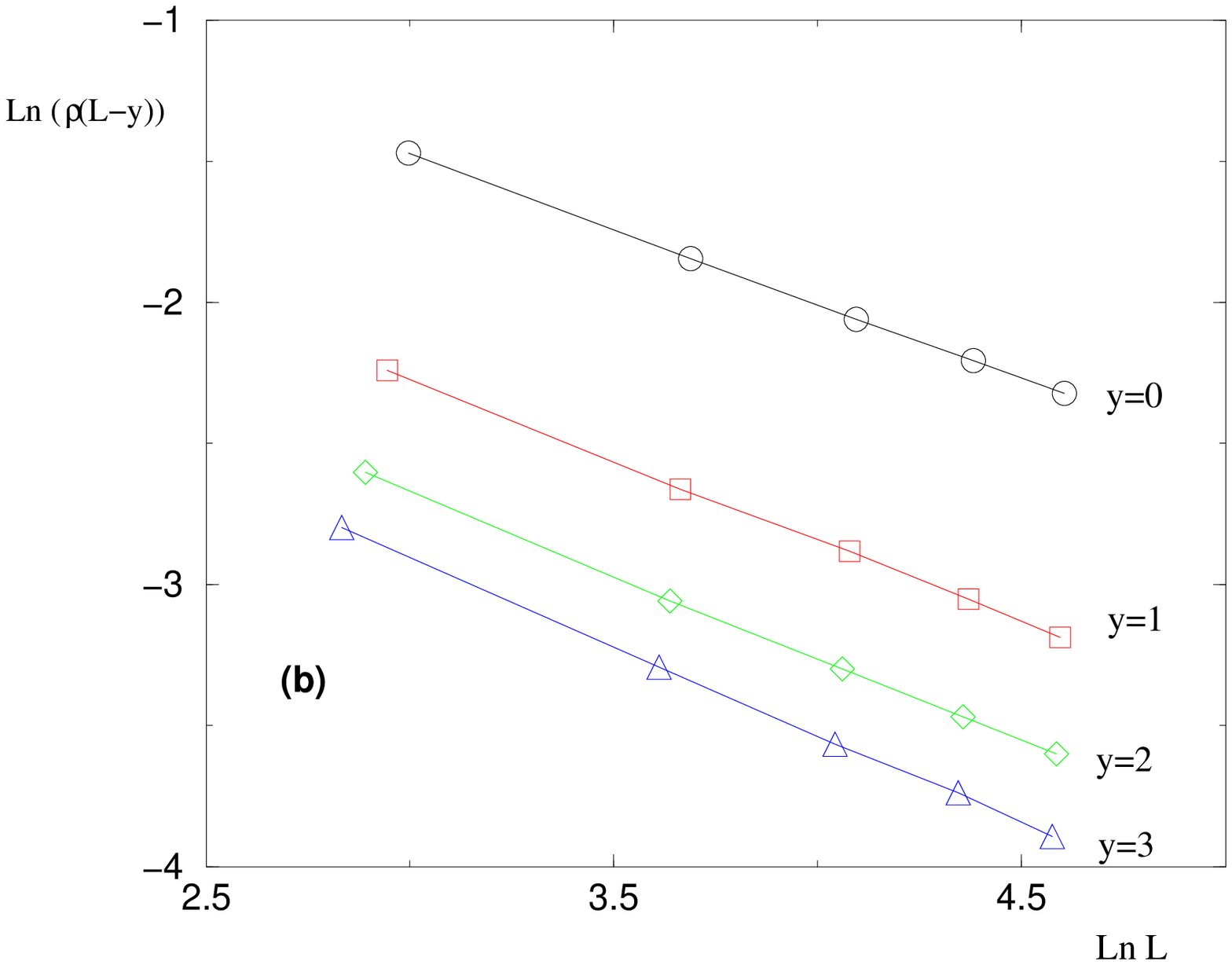}
\caption{$d=2$:(a) Rescaled distribution $R^{boundary}(x=l/L)$
(see equation (\ref{scalingboundary2})), that is singular for $x \to 0$
and $x \to 1$.
(b) Log-log plot of the density $\rho^{boundary}_L(E=0,l=L-y)$
of very large boundary excitations $l=L-y$ with finite $y=0,1,2,3$.}
\label{fig6}
\end{figure}

In $d=2$, we find the same properties for the statistics of boundary
excitations with the appropriate exponent $\theta_2 \sim 0.24$.
The density $\rho^{boundary}_L(E=0,l)$
shown in a log-log plot for various sizes $L$ on Fig. \ref{fig5}(a)
follow the scaling form
\begin{eqnarray}
\rho^{boundary}(E=0,l) = \frac{1}{L^{1+\theta_2}}
 R^{boundary} \left( x = \frac{l}{L} \right) 
\label{scalingboundary2}
\end{eqnarray}
as shown on Fig. \ref{fig5}(b). 
The master curve $R^{boundary}(x)$ shown on Fig. \ref{fig6}(a) 
has the same three important properties as in the $d=1$ case

(i) In the region $x \to 0$, the scaling function follows
the power law
(see the log-log plot on Fig. \ref{fig5}(b))
\begin{eqnarray}
R^{boundary}(x) \oppropto_{x \to 0} \frac{1}{x^{1+\theta_2}}
\end{eqnarray} 
leading to a statistics of independent excitations
\begin{eqnarray}
\rho^{boundary}_L(E=0,l) \sim \frac{1}{l^{1+\theta_2}}
\ \ \ \hbox{for} \ \ \  1 \ll l \ll L
\end{eqnarray}
that follows the droplet power law (\ref{rhodroplet}).

(ii) The function $R^{boundary}(x)$ is minimum at some finite value $x_{min}
(d=2) \sim 0.64 $, and then grows for $x_{min}<x<1$

(iii) The density of
excitations of length $l \sim L$ that branches off at the origin
decays with the power law
 \begin{eqnarray}
\rho^{boundary}_L(E=0,l=L) \sim \frac{1}{L^{\lambda_2}}
\ \ \ \hbox{with} \ \ \  \lambda_2 \sim 0.53
\label{lambda2d}
\end{eqnarray}
as shown on Fig. \ref{fig6}(b).
So the  singularity of the scaling function
$R^{boundary}(x)$ near $x \to 1$ is given by
\begin{eqnarray}
R^{boundary}(x) \oppropto_{x \to 1} \frac{1}{(1-x)^{\sigma_2}}
\ \ \ \hbox{with} \ \ \  \sigma_2 = 1+\theta_2-\lambda_2 \sim 0.71
\end{eqnarray}

\subsection{ Boundary excitations for $d=3$ }

\begin{figure}[htbp]
\includegraphics[height=6cm]{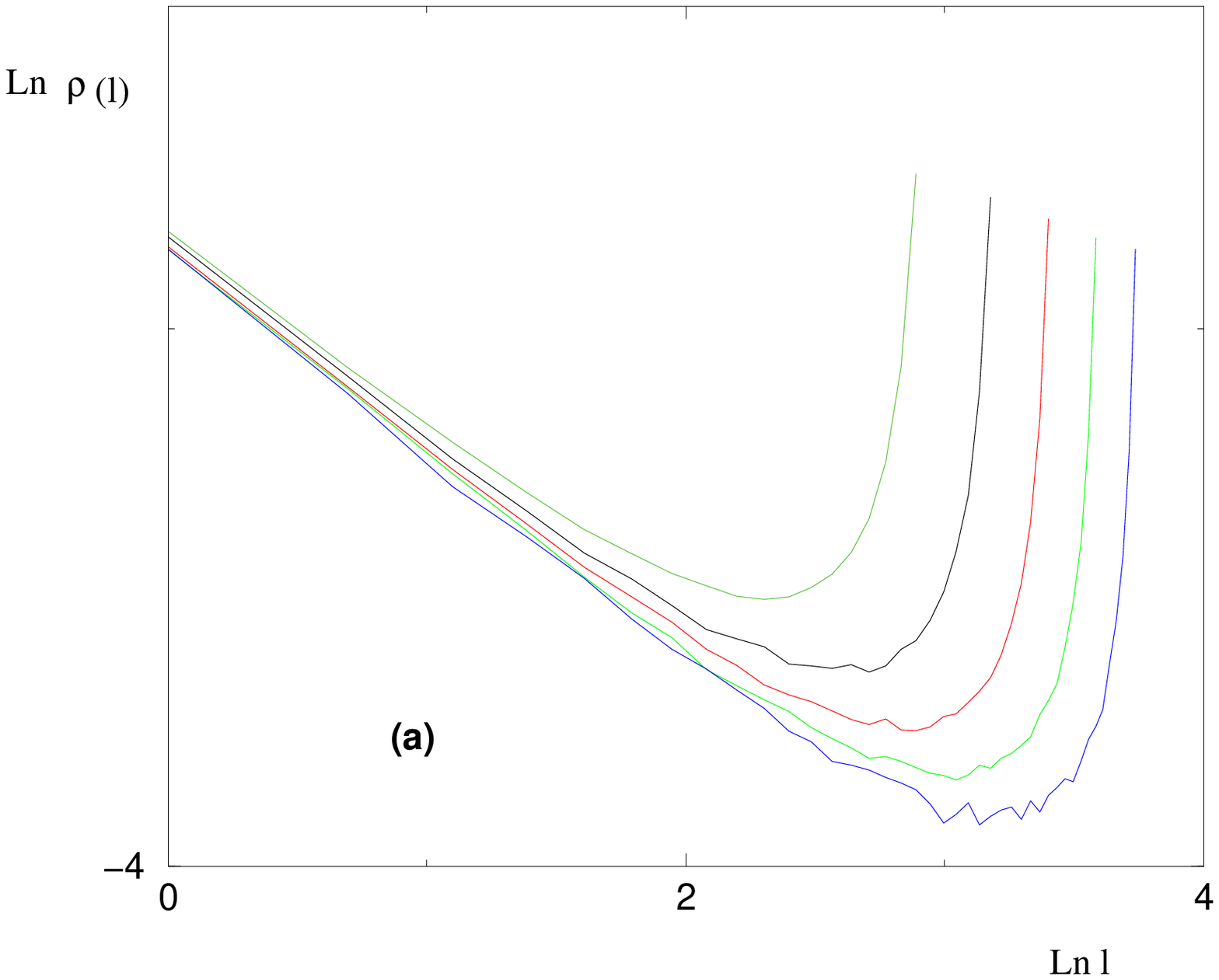}
\hspace{1cm}
\includegraphics[height=6cm]{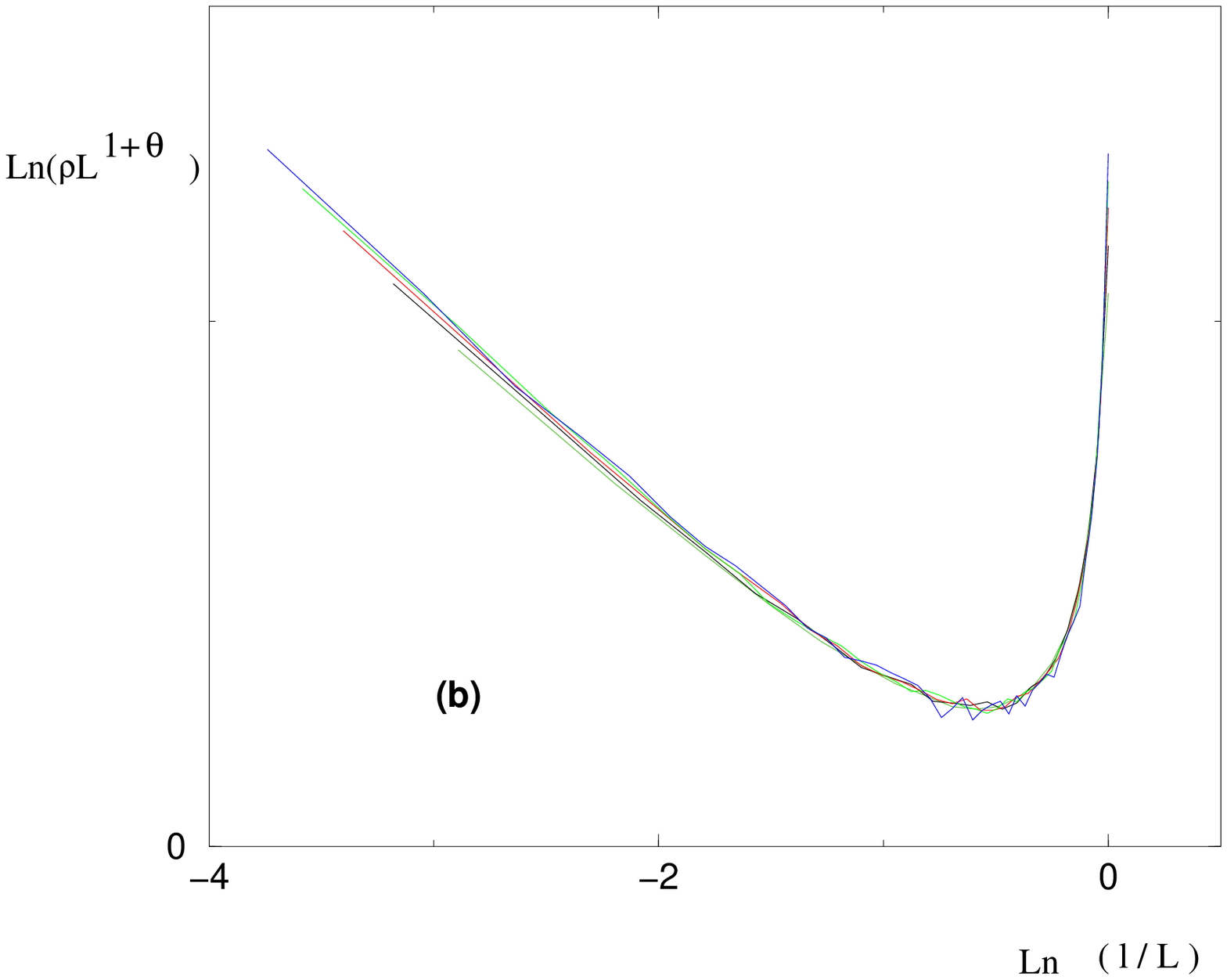}
\caption{(a) Log-log plot of the density $\rho^{boundary}_L(E=0,l)$
of boundary excitations of length $l$ in $d=3$ for
$L=18,24,30,36,42$
(b) Same data after the rescaling of equation (\ref{scalingboundary3})
with $\theta_3 \sim 0.18$.}
\label{fig7}
\end{figure}

\begin{figure}[htbp]
\includegraphics[height=6cm]{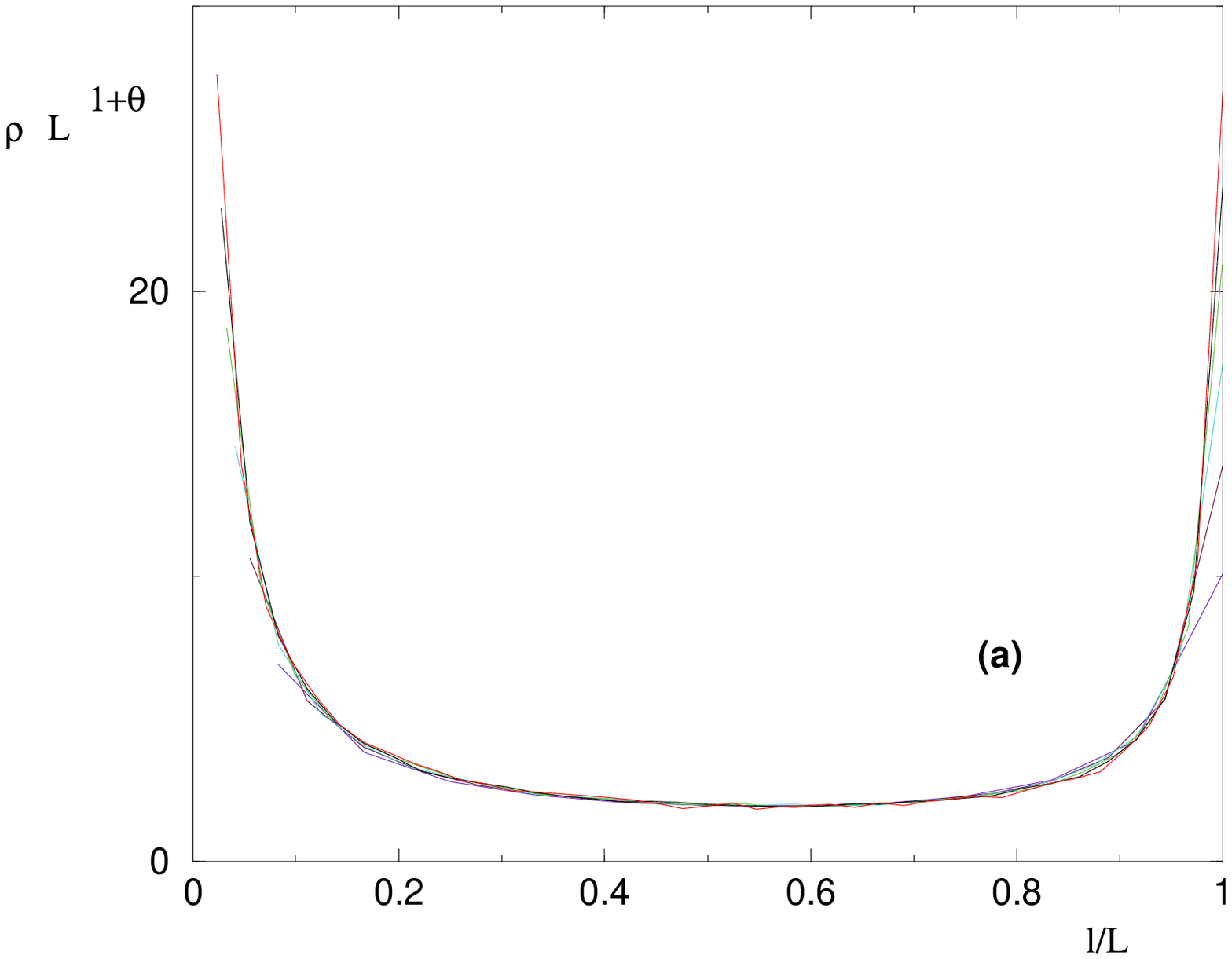}
\hspace{1cm}
\includegraphics[height=6cm]{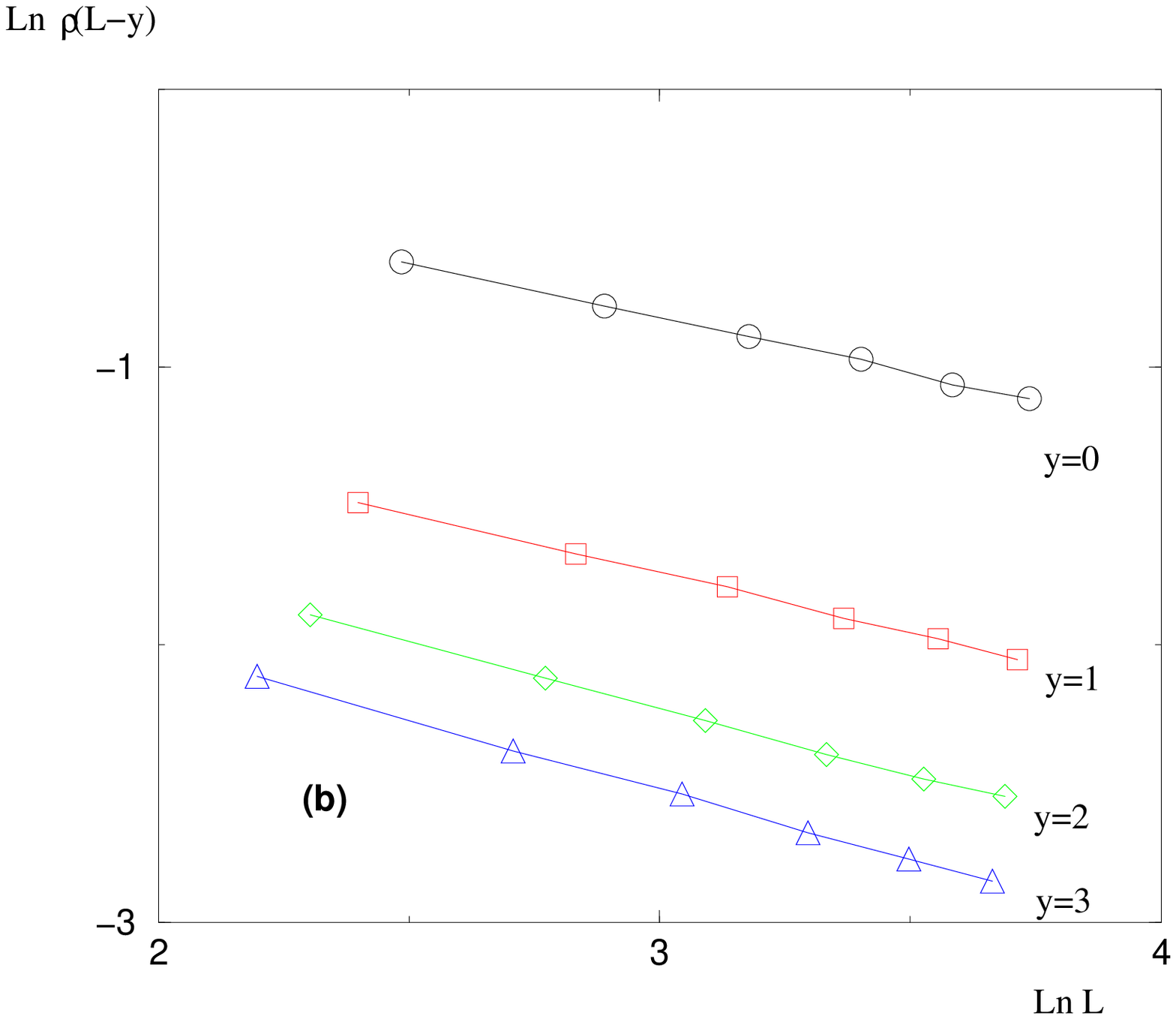}
\caption{$d=3$ :(a) Rescaled distribution $R^{boundary}(x=l/L)$
(see equation (\ref{scalingboundary3})), that is singular for $x \to 0$
and $x \to 1$.
(b) Log-log plot of the density $\rho^{boundary}_L(E=0,l=L-y)$
of very large boundary excitations $l=L-y$ with finite $y=0,1,2,3$.}
\label{fig8}
\end{figure}

In $d=3$, we find again the same properties for the statistics of boundary
excitations with the appropriate exponent $\theta_3 \sim 0.18$.
The density $\rho^{boundary}_L(E=0,l)$
shown in a log-log plot for various sizes $L$ on Fig. \ref{fig7}(a)
follow the scaling form
\begin{eqnarray}
\rho^{boundary}(E=0,l) = \frac{1}{L^{1+\theta_3}}
 R^{boundary} \left( x = \frac{l}{L} \right) 
\label{scalingboundary3}
\end{eqnarray}
as shown on Fig. \ref{fig7}(b). 
The master curve $R^{boundary}(x)$ shown on Fig. \ref{fig8}(a)
has the same three important properties as in the previous cases

(i) In the region $x \to 0$, 
the power law
(see the log-log plot on Fig. \ref{fig7}(b))
\begin{eqnarray}
R^{boundary}(x) \oppropto_{x \to 0} \frac{1}{x^{1+\theta_3}}
\end{eqnarray} 
leads to a statistics of independent excitations
\begin{eqnarray}
\rho^{boundary}_L(E=0,l) \sim \frac{1}{l^{1+\theta_3}}
\ \ \ \hbox{for} \ \ \  1 \ll l \ll L
\end{eqnarray}
that follows the droplet power law (\ref{rhodroplet}).

(ii) The function $R^{boundary}(x)$ is minimum at some finite value $x_{min}
(d=3) \sim 0.56 $, and then grows for $x_{min}<x<1$

(iii) The density of
excitations of length $l \sim L$ that branches off at the origin
decays with the power law
 \begin{eqnarray}
\rho^{boundary}_L(E=0,l=L) \sim \frac{1}{L^{\lambda_3}}
\ \ \ \hbox{with} \ \ \  \lambda_3 \sim 0.39
\label{lambda3d}
\end{eqnarray}
as shown on Fig. \ref{fig8}(b).
So the  singularity of the scaling function
$R^{boundary}(x)$ near $x \to 1$ is given by
\begin{eqnarray}
R^{boundary}(x) \oppropto_{x \to 1} \frac{1}{(1-x)^{\sigma_3}}
\ \ \ \hbox{with} \ \ \  \sigma_3 = 1+\theta_3-\lambda_3 \sim 0.79
\end{eqnarray}

\section{ Statistics of bulk excitations  } 
\label{statexbu}

\subsection{ Measure of independent excitations}

\begin{figure}[htbp]
\includegraphics[height=8cm]{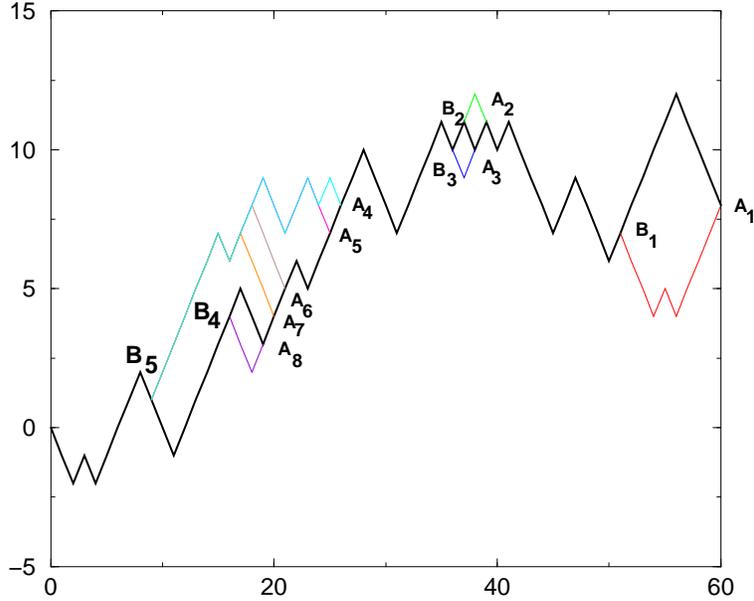}
\caption{ Notion of independent bulk excitations (in $d=1$) :
for each sample, we generate the ground state (bold line)
and consider all points $A_i$ of the ground state,
with partial energy $E_0(A_i)$.
When the energy $E_1(A_i)$
of the second best path ending at $A_i$ satisfies
$E_1(A_i)-E_0(A_i)<T$,
we construct these bulk excitations (light lines) that join again
the ground state at some branch point $B_j$.
These bulk excitations tend to cluster into families.
In our counting procedure of independent excitations,
 two excitations are independent if they have no bond in common.
For instance on the Figure, end-points $A_4$, $A_5$, $A_6$ and $A_7$
count for a single excitation associated to the branch point $B_5$.
The other independent excitations are $(A_1,B_1)$, $(A_2,B_2)$,
$(A_3,B_3)$ and $(A_8,B_4)$.
}
\label{figexbu}
\end{figure}

We now turn to the density $\rho^{bulk}_L(E=0,l)$
of bulk excitations defined as
\begin{eqnarray}
\rho^{bulk}_L(E=0,l) = \lim_{T \to 0} \left[ \frac{l}{T L }
 \overline{ {\cal N}^{bulk}_L(E<T,l) } \right]
\label{defrhobulk}
\end{eqnarray}
where $\overline{{\cal N}^{bulk}_L(E<T,l)}$ is 
now the disorder averaged number of independent bulk excitations
of energy $0<E<T \to 0$ and of length $l$ existing in the bulk
for a directed polymer of length $L$. 
Here the additional normalisation factor $(l/L)$
with respect to the analog definition of
boundary excitations (\ref{defrhoboundary})
ensures a coherent normalization between 
the two densities $\rho^{boundary}$ and $\rho^{bulk}$ :
$\rho^{boundary}(E=0,l)$ represents the probability that the
end-monomer belongs to an excitation of length $l$,
whereas $\rho^{bulk}$ represents the probability that a
bulk-monomer belongs to an excitation of length $l$.
The number ${\cal N}^{indep}_L(E<T,l)$
is measured as follows.
For each sample, we consider all points $\vec r_0(t)$
of the ground state with $t=L,L-1,..,2$ as possible end points of
 bulk excitations.
We consider the best paths joining the ground state at $\vec r_0(t)$
but arriving from different points than $\vec r_0(t-1)$,
to see if they have an relative energy with respect to the ground
state smaller than the cut-off $T$.
If this is the case, we have found a bulk excitation, 
and we measure its length $l$,
i.e. the length over which it is different from the ground state.
Again, as for boundary excitations,
we are interested into independent excitations,
and we use the criterion according to which
two excitations are independent if they have no bond in common.
In $d=1$, since we use polymers with $(\pm)$ steps, this means
that two excitations are independent if they join the ground state
at different points (see Fig. \ref{figexbu}). In $d=2$ and $d=3$,
two excitations are allowed to join the ground state at the same point
provided they branch along two different directions.

\begin{figure}[htbp]
\includegraphics[height=6cm]{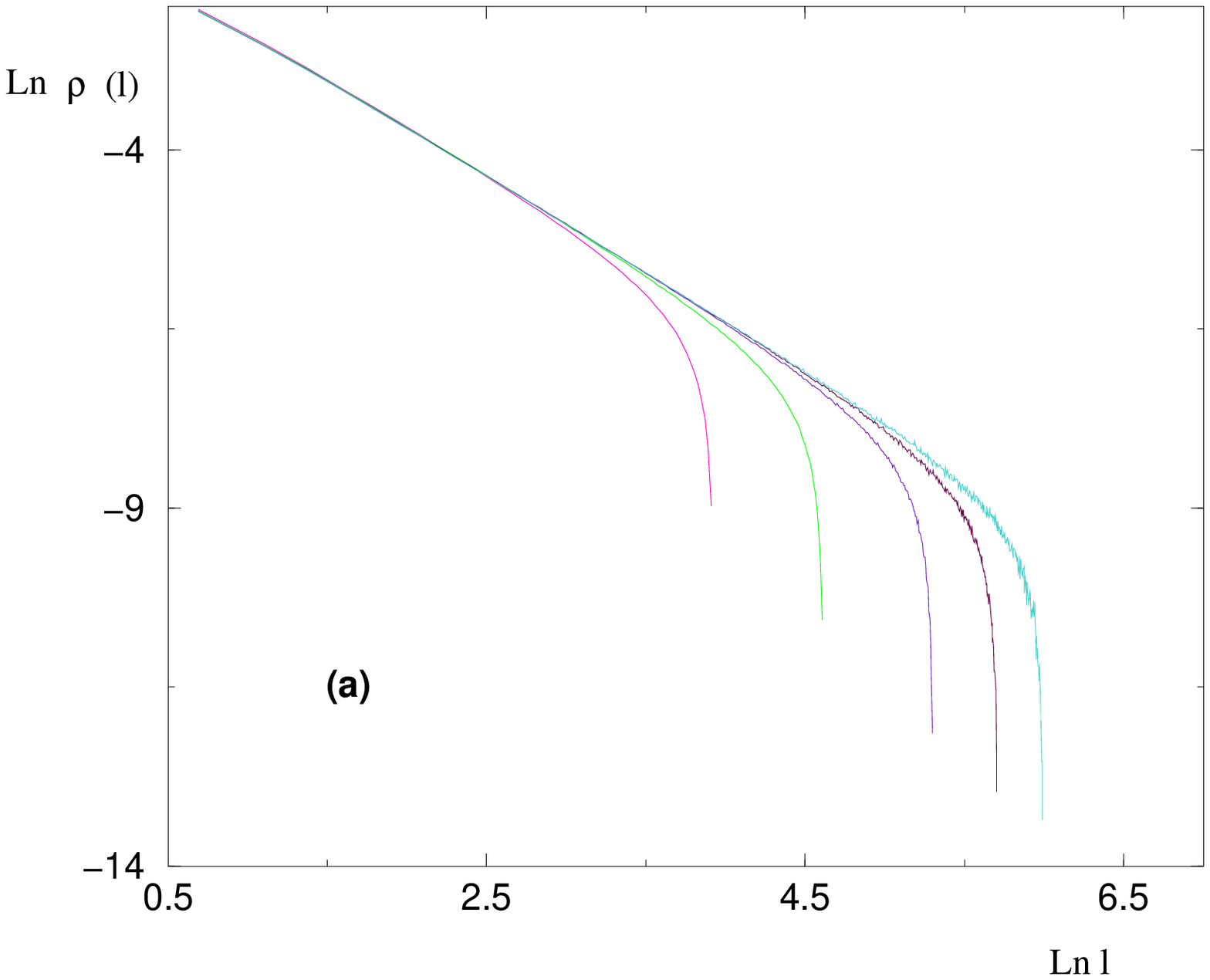}
\hspace{1cm}
\includegraphics[height=6cm]{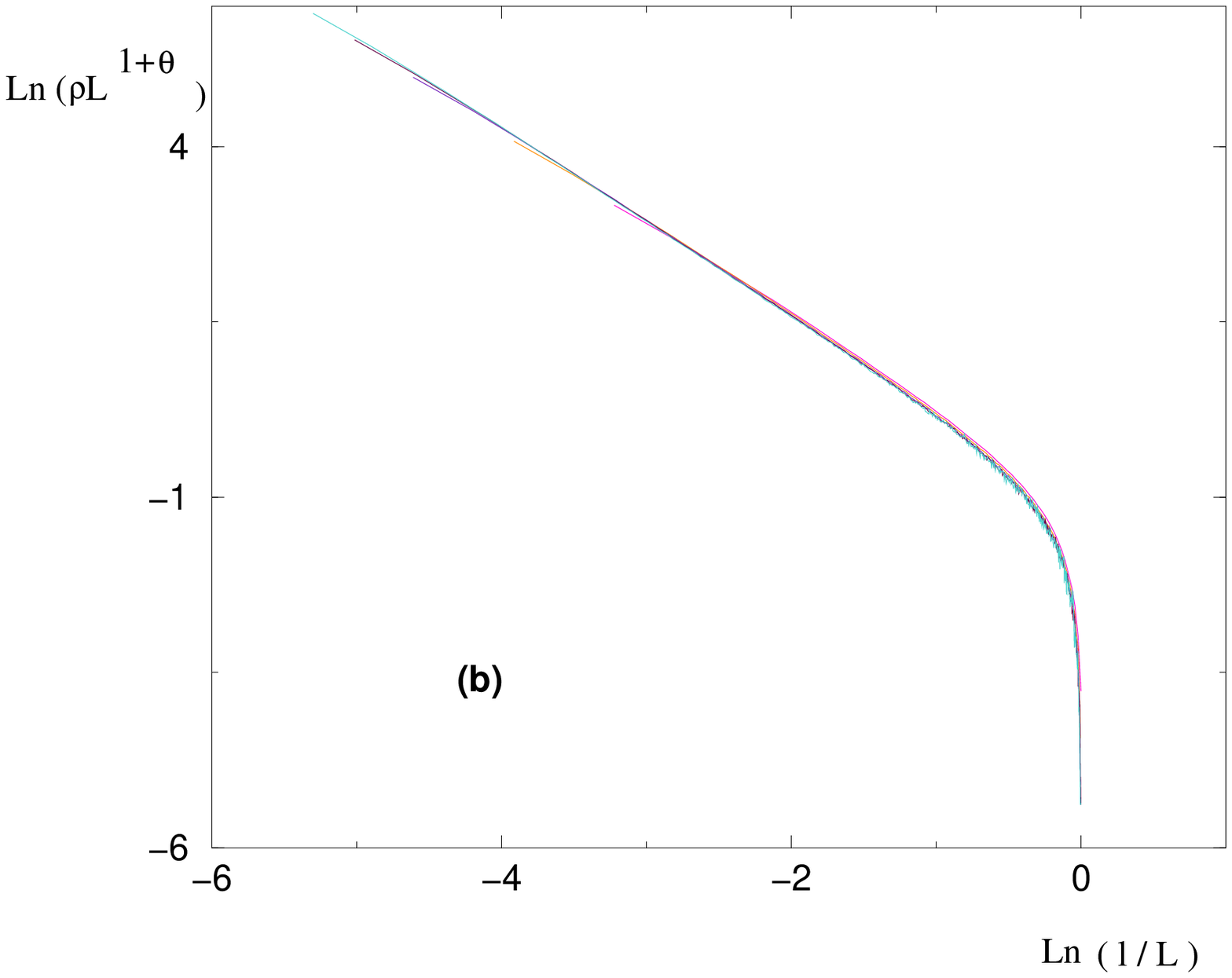}
\caption{(a) Log-log plot of the density $\rho^{bulk}_L(E=0,l)$
of boundary excitations of length $l$ in $d=1$ for
$L=50,100,200,300,400$.
(b) Same data after the rescaling of equation (\ref{scalingbulk})
with exponent $\theta_1=1/3$.}
\label{figbulk1d}
\end{figure}

\begin{figure}[htbp]
\includegraphics[height=6cm]{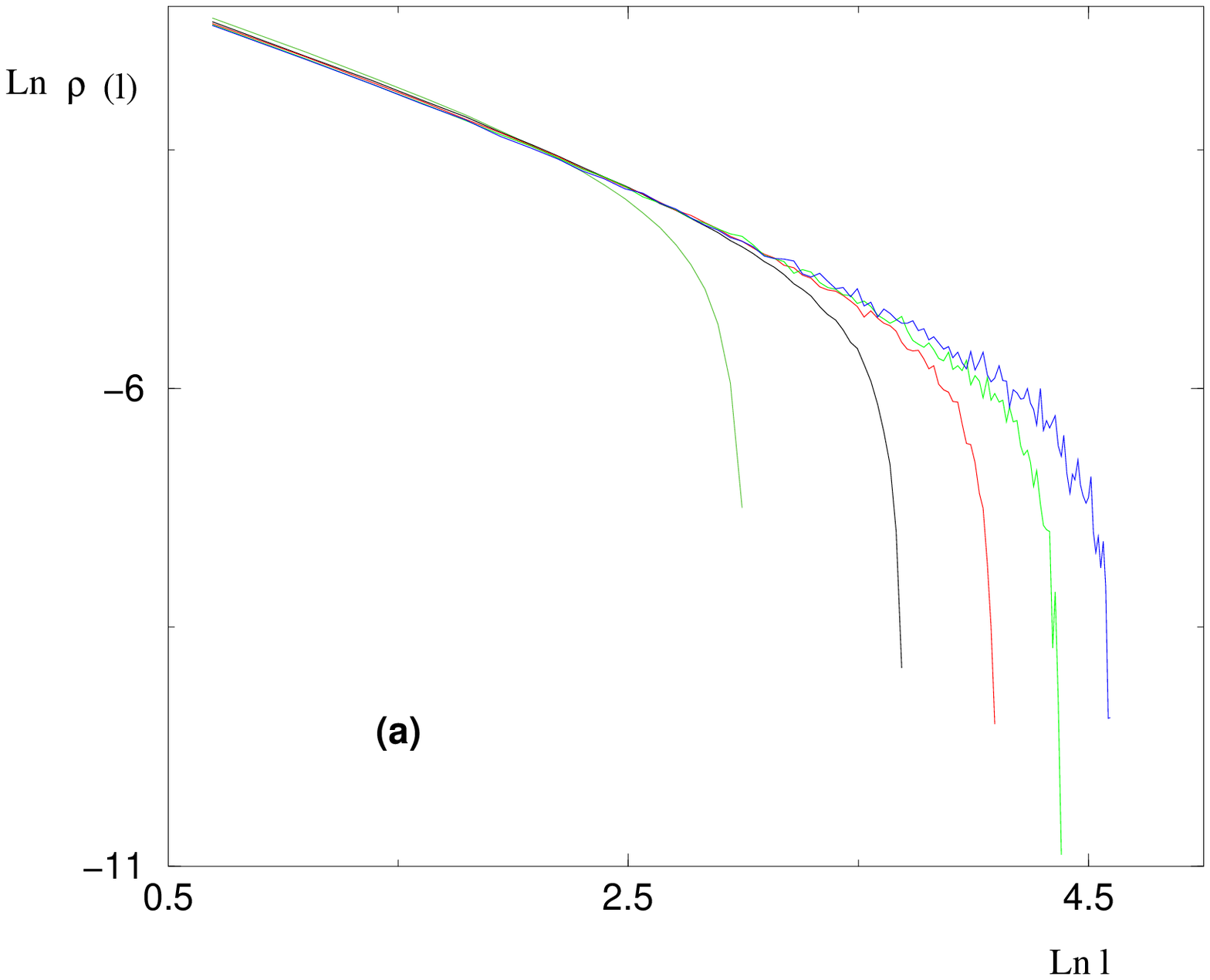}
\hspace{1cm}
\includegraphics[height=6cm]{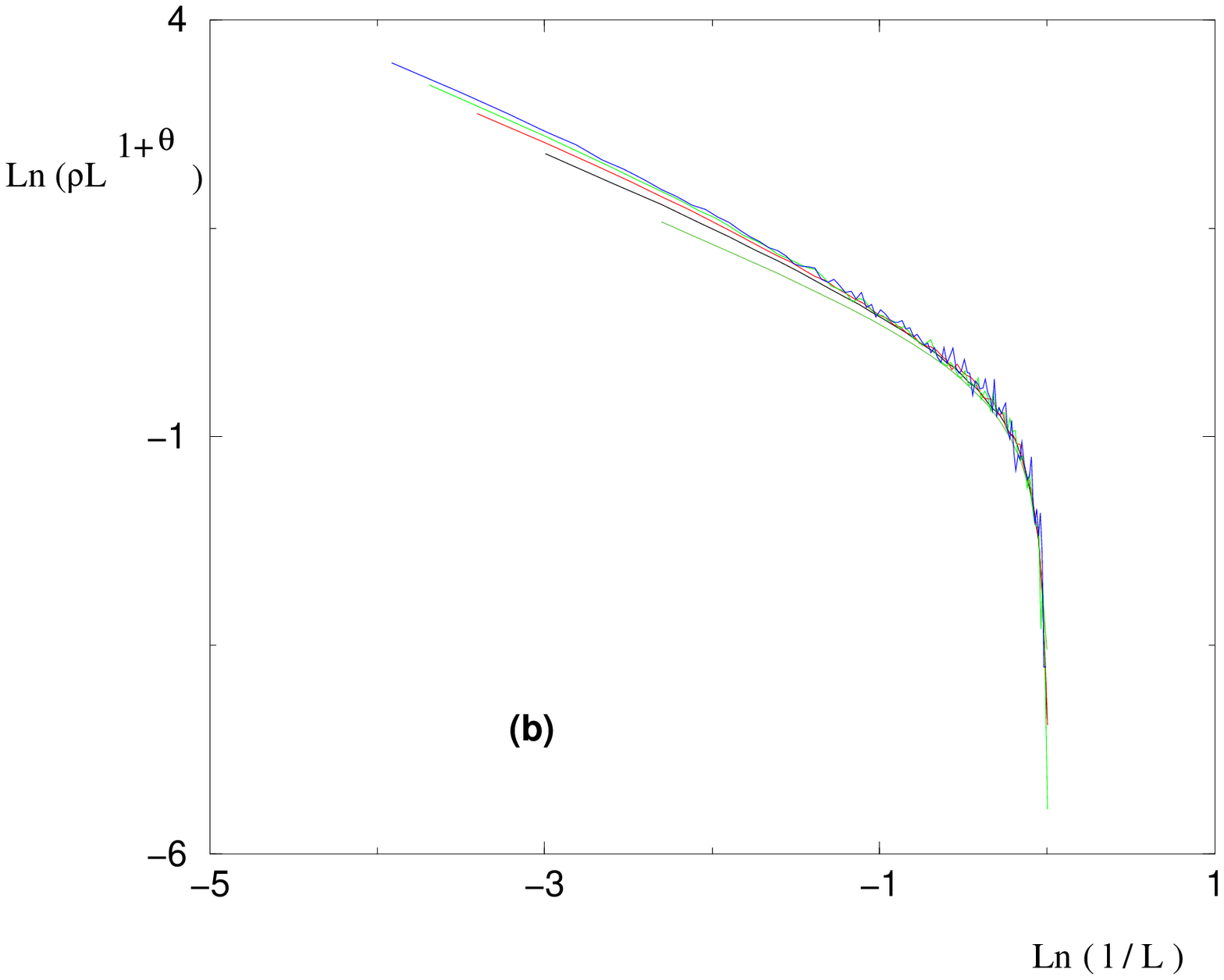}
\caption{(a) Log-log plot of the density $\rho^{bulk}_L(E=0,l)$
of boundary excitations of length $l$ in $d=2$ for
$L=20,40,60,80,100$.
(b) Same data after the rescaling of equation (\ref{scalingbulk})
with exponent $\theta_2=0.24$.}
\label{figbulk2d}
\end{figure}

\begin{figure}[htbp]
\includegraphics[height=6cm]{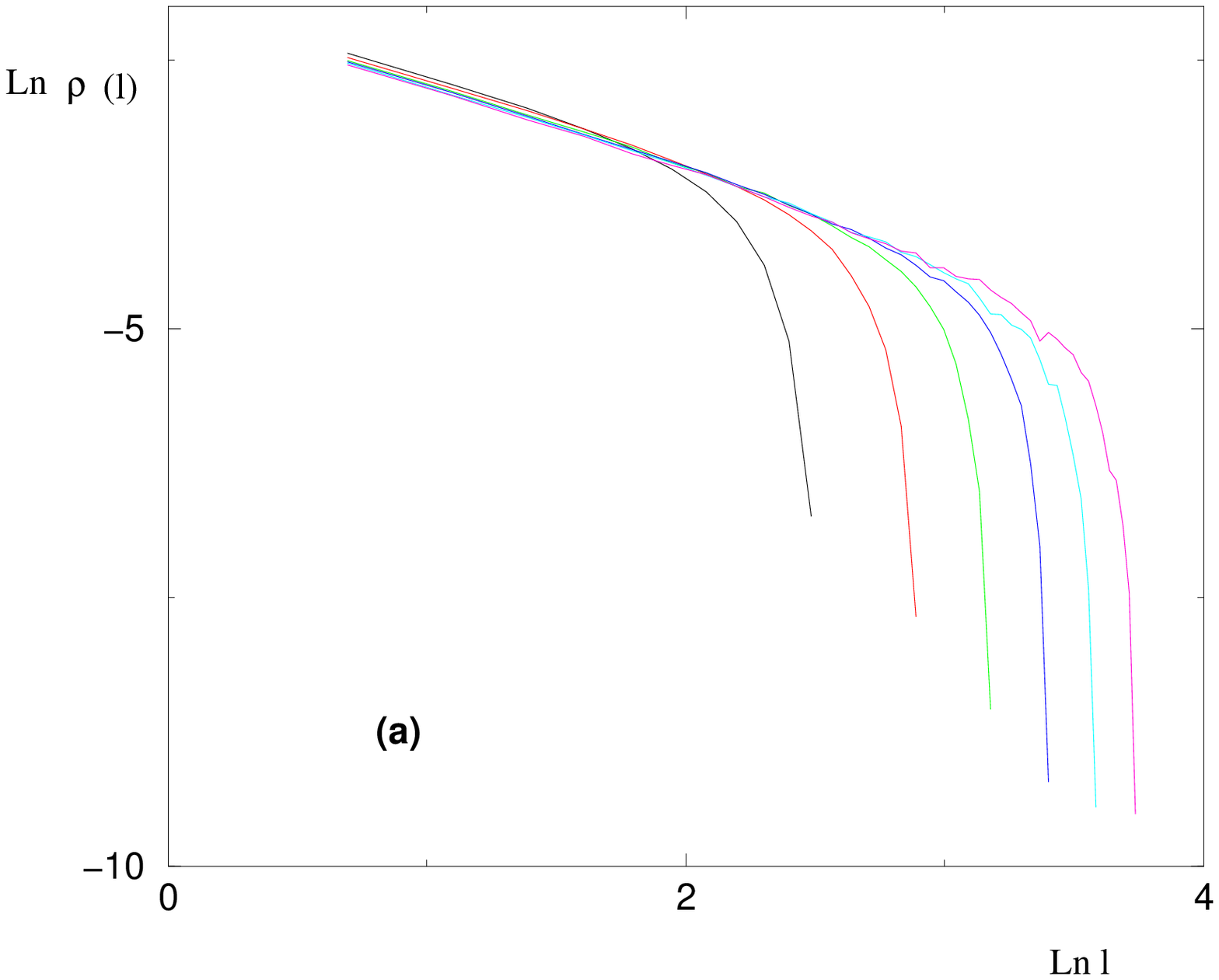}
\hspace{1cm}
\includegraphics[height=6cm]{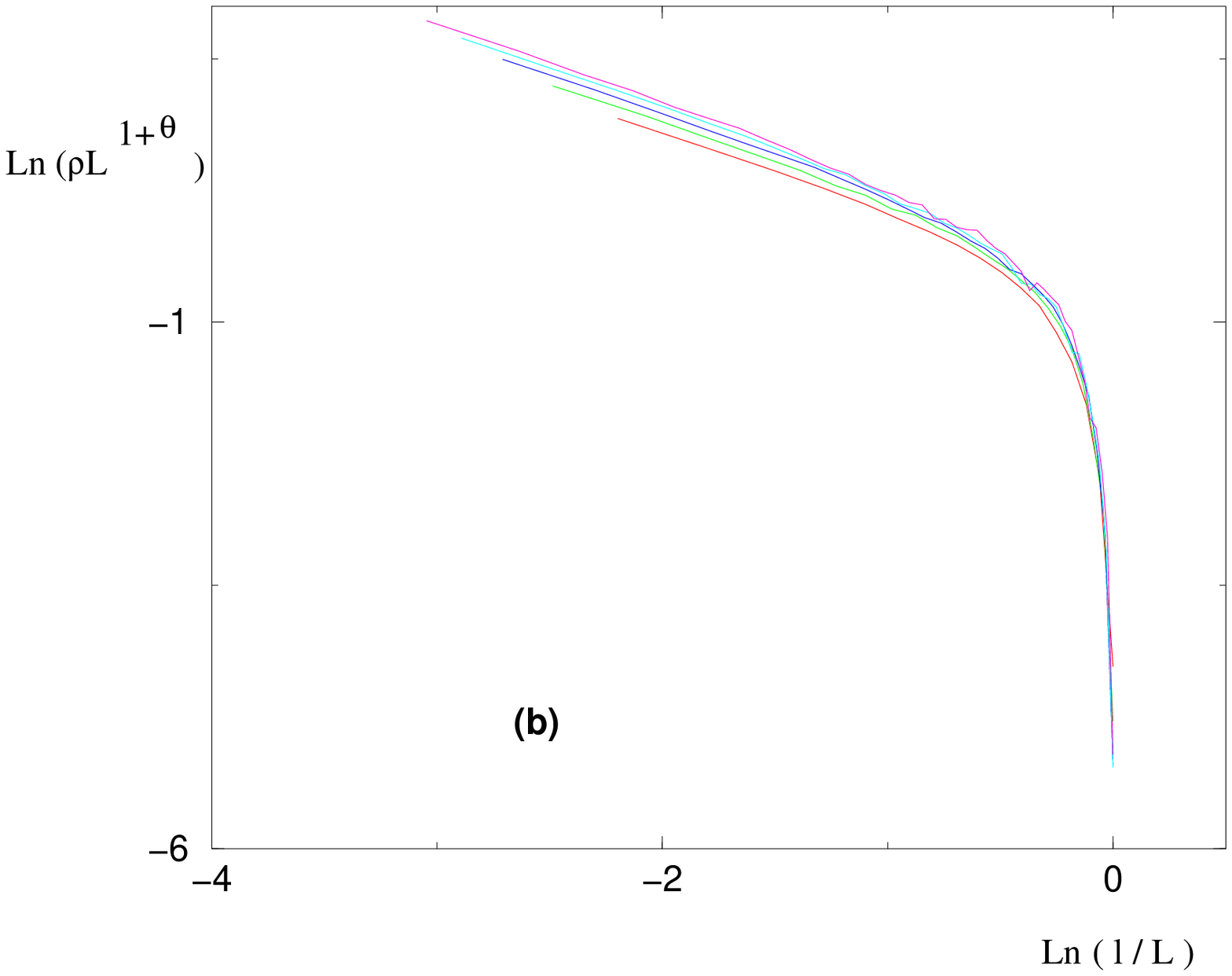}
\caption{(a) Log-log plot of the density $\rho^{bulk}_L(E=0,l)$
of boundary excitations of length $l$ in $d=3$ for
$L=18,24,30,36,42$.
(b) Same data after the rescaling of equation (\ref{scalingbulk})
with exponent $\theta_3=0.18$.}
\label{figbulk3d}
\end{figure}

In dimension $d=1,2,3$, we find that the density of bulk excitations
follows the scaling form 
\begin{eqnarray}
\rho^{bulk}(E=0,l) = \frac{1}{L^{1+\theta_d}}
 R^{bulk} \left( x = \frac{l}{L} \right) 
\label{scalingbulk}
\end{eqnarray}
See Figs \ref{figbulk1d}, \ref{figbulk2d}, \ref{figbulk3d}
for $d=1,2,3$ respectively. As $d$ increases, the quality of the
rescaling gets weaker, because of the smaller sizes $L$ that can be
studied via transfer matrix.
 
In contrast with the scaling function $R^{boundary}(x)$
of boundary excitations, the scaling function $R^{bulk}(x)$
decays monotonically for $0<x<1$.
In the region $x \to 0$, 
the scaling function follows the power law
\begin{eqnarray}
R^{bulk}(x) \oppropto_{x \to 0} \frac{1}{x^{1+\theta_d}}
\end{eqnarray} 
so that in the regime $ 1 \ll l \ll L$, 
 the statistics of independent excitations
\begin{eqnarray}
\rho^{bulk}_L(E=0,l) \sim \frac{1}{l^{1+\theta_d}}
\ \ \ \hbox{for} \ \ \  1 \ll l \ll L
\end{eqnarray}
follows the droplet power law (\ref{rhodroplet}).

\section{Summary and Conclusions}
\label{conclusion}
\begin{figure}[htbp]
\includegraphics[height=6cm]{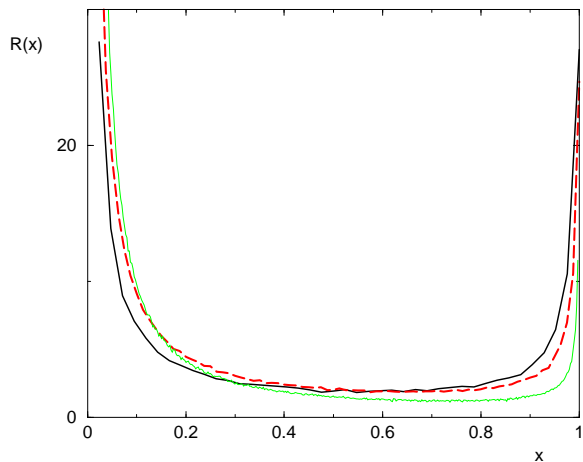}
\caption{Comparison of the scaling function  $R^{boundary}(x)$ for
$d=1$ (thin line), $d=2$ (dashed line) and $d=3$ (thick line).}
\label{figboundary}
\end{figure}

We have studied the low energy excitations of a directed polymer 
in a $1+d$ random medium. For dimensions $d=1,2,3$,
we find that the densities of bulk and boundary excitations follow the
scaling behavior $\rho^{bulk,boundary}_L(E=0,l) = L^{-1-\theta_d}
R^{bulk,boundary}(x=l/L)$.
In the limit $x=l/L \to 0$, both scaling functions
$R^{bulk}(x)$ and $R^{boundary}(x)$ behave as
$R^{bulk,boundary}(x) \sim x^{-1-\theta_d}$, leading to the droplet
power law $\rho^{bulk,boundary}_L(E=0,l)\sim l^{-1-\theta_d} $ in the
regime $1 \ll l \ll L$. Beyond
their common singularity near $x \to 0$, the two scaling functions 
$R^{bulk,boundary}(x)$ are very different 
(this shows the importance of boundary conditions) :
whereas $R^{bulk}(x)$ decays monotonically for $0<x<1$, the function
$R^{boundary}(x)$ first decays for $0<x<x_{min}$, then grows for
$x_{min}<x<1$, and finally presents a power law singularity
$R^{boundary}(x)\sim (1-x)^{-\sigma_d}$ near $x \to 1$.
The density of excitations of length $l \sim L$
decays as $\rho^{boundary}_L(E=0,l=L)\sim L^{- \lambda_d} $
where $\lambda_d=1+\theta_d-\sigma_d$. Our numerical estimates
$\lambda_1 \simeq 0.67$, $\lambda_2 \simeq 0.53$ and $\lambda_3 \simeq
0.39$  suggest the relation $\lambda_d= 2 \theta_d$, although
we are not aware of any simple argument to justify it.
However, if it holds, this would mean that the scaling function
$R^{boundary}(x)$ has singularities with exponents $(1+\theta_d)$
and $(1-\theta_d)$ near $x \to 0$ and $x \to 1$ respectively,
i.e. these singularities tend to become the same
as $\theta_d$ decreases, i.e. as the dimension $d$ increases
(see Fig. \ref{figboundary} ). This trend is reminiscent of the result
on the Cayley tree discussed in Eq. (\ref{overlaptang},\ref{overlaptangsing})
where the singularities have the same exponent on both sides,
even if the value $(3/2)$ seems specific to the tree structure
and cannot be obtained as the limit $\theta_d=0$ in our results.

Let us now mention how one recovers the identity (\ref{cumulant})
of the statistical tilt symmetry. A boundary excitation of length $l$
is expected to give rise to a fluctuation of the end-point of order
$\Delta r \sim l^{\zeta}$ (we temporarily drop the dimension
dependence of the exponents), so that at order $T$, one gets
\begin{eqnarray}
\overline{ <\Delta r^2>} \sim T \int_1^L dl \ \ l^{2 \zeta} \rho_L^{boundary}(E=0,l)
\end{eqnarray}
The contribution of excitations of length $l$ with $x=l/L$ finite
reads
\begin{eqnarray}
\left[ \overline{ <\Delta r^2>}  \right]_{0<x<1} =
 T L^{2 \zeta -\theta} \int_0^1 dx \ \ x^{2 \zeta} R^{boundary}(x)
\end{eqnarray}
whereas the contribution of very large excitations of length $l=L-y$ with
finite $y$ reads
\begin{eqnarray}
\left[\overline{ <\Delta r^2>}\right]_{ x \sim 1} \sim T L^{2 \zeta - \lambda }
\end{eqnarray}
Since $\lambda > \theta$, the leading contribution is the first one.
Using the scaling relation (\ref{zetaomega}), this contribution is of
order $L^{2 \zeta -\theta}=L$, as it should to recover (\ref{cumulant}). 

Finally, since the directed polymer model plays the role of a ` baby
spin glass', and since various numerical studies on spin glasses
\cite{numerical} find both the droplet scaling behavior for small
excitations and system-size excitations governed by another `global'
exponent, one may wonder whether both types of excitations can be
understood within a single scaling function $R(x)$ of the volume
fraction $x=v/V$, where the droplet exponent describes the power law in
the regime $x \to 0$, whereas the statistics of system-size
excitations depends on the global properties of the scaling function
$R(x)$.

\end{document}